# Spontaneous and Stimulated Radiative emission of Modulated Free-Electron Quantum wavepackets - Semiclassical Analysis

Yiming Pan, Avraham Gover

*Department of Electrical Engineering Physical Electronics,*

*Tel Aviv University, Ramat Aviv 69978, ISRAEL*

**Abstract**

Here we present a semiclassical analysis of spontaneous and stimulated radiative emission from unmodulated and optically-modulated electron quantum wavepackets. We show that the radiative emission/absorption and corresponding deceleration/acceleration of the wavepackets depend on the controllable 'history-dependent' wavepacket size. The characteristics of the radiative interaction when the wavepacket size (duration) is short relative to the radiation wavelength, are close to the predictions of the classical point-particle modeling. On the other hand, in the long-sized wavepacket limit, the interaction is quantum-mechanical, and it diminishes exponentially at high frequency. We exemplify these effects through the scheme of Smith-Purcell radiation, and demonstrate that if the wavepacket is optically-modulated and periodically-bunched, it exhibits finite radiative emission at harmonics of the modulation frequency beyond the limit of high-frequency cutoff. Besides, the radiation analysis is further extended to the cases of superradiant emission from a beam of phase-correlated modulated electron wavepackets. The features of the wavepacket-dependent radiative emission explain the classical-to-quantum theory transition, and indicate a way for measuring the quantum electron wavepacket size. This suggests a new direction for exploring light-matter interaction.

**Introduction**

Accelerated free electrons emit electromagnetic radiation when subjected to an external force (e.g. synchrotron radiation [1], Undulator radiation [2], Compton scattering [3]). Radiation can also be emitted by currents that are induced by free electrons in polarizable structures and materials, such as in Cherenkov radiation [4], transition radiation [5], Smith-Purcell radiation [6]. Some of these schemes were demonstrated to operate as coherent stimulated radiative emission sources, such as Free Electron Lasers (FEL) [7-9], as well as accelerating (stimulated absorption) devices, such as Dielectric Laser Accelerator (DLA) and Inverse Smith-Purcell effect [10-12].

The stimulated radiative emission of an ensemble of electrons (an electron beam) is coherent (to the extent of coherence of the input radiation wave being amplified). The spontaneous emission of an electron beam in any of these radiation schemes is incoherent, unless the particles are made to emit in phase with each other. This can be done by pre-bunching the beam. In this case, the radiative emission is proportional to $N^2$ - the number of electrons squared (while the emission of a randomly distributed electron beam is proportional to N). This coherent spontaneous radiation process is analogous to Dicke's superradiance of atomic dipoles [13]. It has been extended in the classical limit to the case of bunched electron beams, employing a general formulation that is applicable to the wide variety of the aforementioned free electron radiation schemes. [14].

Most of the free electron radiation schemes of emission or acceleration operate in the classical theoretical regime of electrodynamics, where the electrons can be considered point-particles and the radiation field is described by Maxwell equations (no field quantization). However a variety of free electron radiation schemes [15,16], and particularly FEL [e.g. Refs: 17,18,19] have been analyzed in the framework of a quantum model in which the electron is described in the inherently quantum limit - given as a plane-wave quantum wave function – the opposite limit of the point-particle classical presentation. Quantum description of the electron wavefunction is also used in another recently developed research field of electron interaction with radiation: Photo-Induced Near-Field Electron Microscopy (PINEM) [20,21] In this scheme a single electron quantum wavefunction interacts with the near-field of a nanometric structure illuminated by a coherent laser beam. Of special relevance for the present discussion is a recent PINEM-kind experiment of Feist et al [22], in which it was demonstrated that optical frequency modulation of the energy and density expectation values of a single electron wavepacket are possible in this method.

All these theoretical models and experiments in the classical and quantum limits of the electron description raise interest in the theoretical understanding of the transition of the electron-radiation interaction process from the quantum to the classical limit. This is also related to deeper understanding of fundamental physics questions, such as the particle-wave duality nature of the electron, [23] and the interpretation and measurability of the electron quantum wavepacket.

The wavepacket regime of electron interaction with radiation is not well founded in theory. Recent experimental study of spontaneous Compton scattering by the expanding wavepacket of a single electron, revealed no dependence on the wavepacket size and history, as also was predicted by a theoretical QED analysis of this problem [24-26]. We assert, though, that this conclusion does not carry over to the case of stimulated interaction (emission/absorption or acceleration/deceleration). We have shown in an earlier publication [27] that the classical phase-dependent acceleration/deceleration of a single electron in the point-particle limit is valid in a certain operating range also in the quantum-wavepacket regime. The momentum transfer from the field to the wavepacket (acceleration) is smaller than in the point-particle limit, and it diminishes in the inherent quantum limit, where the wavepacket size $\sigma_t$ exceeds the optical radiation period $T = 2\pi/\omega$ of the interacting radiation wave

$$\omega\sigma_t > 1 \qquad (1)$$

Thus, measurements of the electrons energy spectrum after interaction with radiation waves at different frequencies would enable determination of the history-dependent wavepacket size.

In the semi-classical analysis of electron wavepacket interaction with radiation [27] the wavepacket-dependent energy (momentum) acceleration/deceleration of the electron was calculated using first order analysis and exact numerical solution of Schrodinger equation. Evidently, such a change in the electron wavepacket energy must involve also corresponding change in the energy of the interacting radiation wave. In the present article, we examine this energy exchange process on the side of the radiation field, again using a semi-classical formulation, in which the electron current density is represented by the expectation value of the wavepacket probability density distribution, and the radiation field is classical, and modeled in terms of a modal expansion formulation of classical Maxwell equations. The following analysis results in full agreement with the earlier Schrodinger semi-classical analysis and presents the same distinction between the quantum, classical and wavepacket interaction regimes. Further, here we also present for the first time expressions for the spontaneous and stimulated emission from a modulated electron wavepacket and from an ensemble of unmodulated or modulated electron wavepackets.

In the following sections, we present a detailed semi-classical theory for wavepacket-dependent radiation of single electron and electron beams. In section "Modeling and Methods", we derive the probability density current of an unmodulated and modulated electron quantum wavefunction from its Schrodinger Equation solution. We then use these current expressions to find the spectral optical parameters of the emitted radiation based on a general mode-expansion formulation solution of Maxwell equations. The results and discussions are exhibited in the following sections "Results and Discussions", which are divided into four cases of spontaneous and stimulated emissions: In subsections I.A and I. C respectively we analyze the cases of unmodulated and modulated single electron wavepacket; in subsections IIE-IIG we analyze the cases of superradiant and stimulated-superradiant radiation emission by a multiple-particle electron beam of unmodulated wavepackets and modulated wavepackets. In subsection I.B we derive a classical "Einstein relation" between spontaneous emission and stimulated emission/absorption of an electron wavepacket. In subsection I.D we present a detailed example of wavepacket radiative emission/interaction in a Smith-Purcell radiation scheme. Finally, in the last section "Conclusions and Outlook" we summarize all new results, and propose an experimental setup for testing the dependence of wavepacket radiative emission/interaction on the wavepacket characteristics.

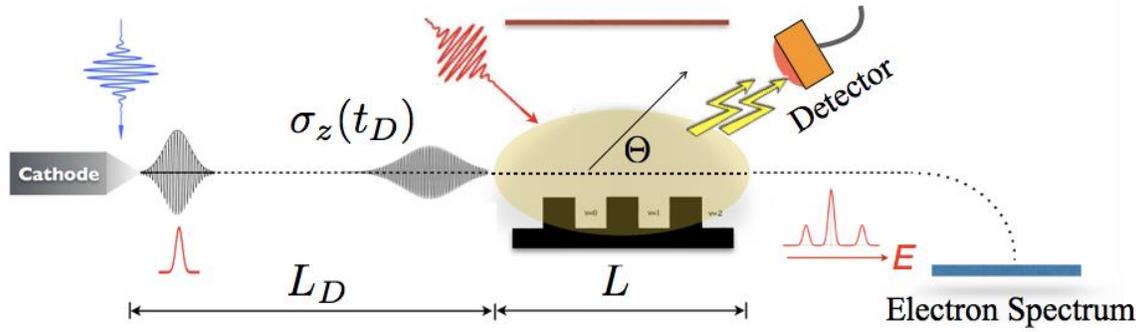

*Figure 1: Experimental setup of wavepacket-dependent spontaneous/stimulated Smith-Purcell radiation emission/absorption and corresponding deceleration/acceleration of quantum electron wavepacket.*

**Modeling and Methods**

Here, we present a semi-classical analysis of spontaneous, superradiant and stimulated superradiant emission by modulated and unmodulated electron quantum wavepackets and multi-particle beams.

A proposed scheme for measuring spontaneous and stimulated radiation emission and electron energy spectrum of an electron wavepacket is shown in Fig. 1. This interaction scheme, based on the Smith-Purcell radiation effect was used in [27] to calculate the wavepacket-dependent electron energy spectrum due to radiative interaction with an input radiation field, injected into the interaction region above the grating, in controlled phase correlation with the incoming electron wavepacket. The wavepacket size depends on the drift time from the cathode to the grating. Here we include also optical light detection for measuring the spontaneous and stimulated emission from the electron wavepacket.

Fig. 2 shows schematically an elaboration of the first scheme, including an energy modulation region where the electron wavepacket traverses the near field region of a tip illuminated by a laser tip [22], and gets energy-modulated at the frequency $\omega_b$ of the "modulating radiation wave". The energy modulation turns into density modulation of the wavepacket envelope within the drift length $L_D' = L_d - L_c$. Then, in the interaction region 0<z<L_G above the grating, it interacts with the near-field of an "input radiation wave" at a frequency $\omega_0$ near the frequency of the "modulating wave" or its harmonic frequency - $\omega_0 \simeq l\omega_b$. Under the force field of the "input wave", the modulated electron wavepacket experiences acceleration/deceleration, and exhibits a corresponding stimulated radiation-emission/absorption, depending on the phase difference between the "input wave" and the "modulating wave". According to classical electrodynamics analysis, the electron wavepacket can emit spontaneously radiation, also when the "input wave" is turned "off", at harmonics of the modulation frequency, beyond the frequency cut-off condition (1) of an unmodulated wavepacket.

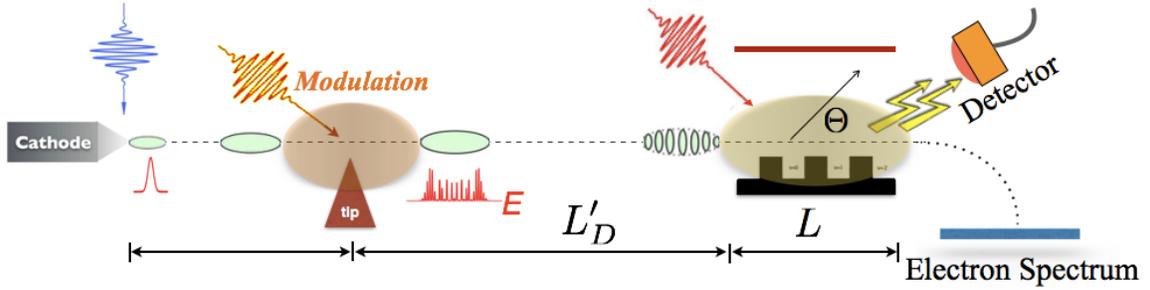

*Figure 2: Experimental setup of wavepacket-dependent Stimulated-Superradiant Smith-Purcell radiation emission/absorption and deceleration/acceleration with a harmonic of a density-bunched quantum electron wavepacket. The wavepacket is energy modulated at a tip by multi-photon emission/absorption process, and turns to be density modulated after a drift length $L_D$'.*

**Semi-classical derivation of the probability density current of an electron quantum wavepacket**

We model the electron wavefunction after its emission from the cathode by a Gaussian wavepacket in momentum space

$$\psi(z,t) = \frac{1}{\sqrt{2\pi\hbar}} \int dp \, \psi_p e^{i(pz-E_p t)/\hbar} \qquad (2)$$

$$\psi_p = \left(2\pi\sigma_{p_0}^2\right)^{-1/4} \exp\left[-\frac{(p-p_0)^2}{4\sigma_{p0}^2}\right] \tag{3}$$

To evaluate the evolution of the wavefunction (2) in time and space we use the Taylor expansion of the relativistic electron dispersion relation

$$E_p = c\sqrt{m_0^2 c^2 + p^2} \approx E_{p_0} + v_0(p-p_0) + \frac{(p-p_0)^2}{2m^*} \tag{4}$$

where $E_{p_0} = \gamma_0 m_0 c^2$, $p_0 = \gamma_0 m v_0$, $m^* = \gamma_0^3 m$ are the center energy and momentum and the effective longitudinal mass of the wavepacket, and $\gamma_0 = 1/\sqrt{1-v_0^2/c^2}$. Substitution of (4) and (3) in (2) results in [15,27]

$$\psi(z,t) = \frac{e^{i(p_0 z - E_{p_0} t)/\hbar}}{\left(2\pi\sigma_{z_0}^2\right)^{1/4}\sqrt{1+i\xi t}} \exp\left[\frac{(z-v_0 t)^2}{4\sigma_{z_0}^2(1+i\xi t)}\right] \tag{5}$$

with $\xi = \frac{\hbar}{2m^*\sigma_{z_0}^2}$ and the initial wavepacket "waist" $\sigma_{z_0} = \hbar/2\sigma_{p_0}$.

The expectation value of the free drifting electron current density can be written in terms of the expectation value of the electron probability density

$$\mathbf{J}(\mathbf{r},t) = -\frac{e\hbar}{2m^*}\langle\psi^*\nabla\psi - \psi\nabla\psi^*\rangle \simeq -e\mathbf{v}_0|\psi(\mathbf{r},t)|^2 \tag{6}$$

In our 1-D model, the axial current is

$$J_0(z,t) = -ev_0|\Psi_0(z,t)|^2 = -ev_0 f_e(\mathbf{r}_\perp) f_{ez}(z-v_0 t) \tag{7}$$

where

$$f_e(z-v_0 t) = \frac{1}{\sqrt{2\pi}\sigma_z(t)} \exp\left[-\frac{(z-v_0 t)^2}{2\sigma_z^2(t)}\right]$$

$$\sigma_z(t) = \sigma_{z_0}\sqrt{1+\xi^2 t^2} \tag{8}$$

and the transverse expansion of the transverse profile function $f_e(\mathbf{r}_\perp)$ is neglected.

Now transform the coordinates frame, so that the origin $z = 0$ is at the entrance to the interaction region, and the electron wavepacket arrives there at time $t_{0e}$ after drift time $t_D$, and further assume that the electron wavepacket dimensions hardly change along the interaction length: $\sigma_z = \sigma_z(t_D)$, or correspondingly $\sigma_t = \sigma_t(L_D)$, then we can write

$$J_0(z,t) = -ef_{e\perp}(\mathbf{r}_\perp)f_e(t - t_{0e} - z/v_0)$$
$$f_e(t) = \frac{1}{\sqrt{2\pi}\sigma_t(L_d)} e^{t^2/2\sigma_t^2(L_D)} \qquad (9)$$
$$\sigma_t(L_D) = \sigma_{t0}\sqrt{1 + (\xi/v_0)^2 L_d^2}$$

Further elaboration is required for describing the wavepacket evolution at the modulation region in the case shown in Fig. 2 and the subsequent drift hereafter (see Appendix A). In this case, the electron wavepacket undergoes a multi photon emission/absorption process in the short near-field modulation region, and its quantum wavefunction gets modulated in momentum space at harmonics of the photon recoil momentum $\delta p = \hbar\omega_b / v_0$, where $\omega_b$ is the frequency of the modulating laser. After the modulation point

$$\psi_p = \left(2\pi\sigma_{p_0}^2\right)^{-1/4} \sum_{n=-\infty}^{\infty} J_n(2|g|)\exp\left(-\frac{(p - p_0 - n\delta p)^2}{4\sigma_{p_0}^2}\right) \qquad (10)$$

where $2|g| \approx (e/\hbar\omega_b)\int_0^{L_M} dz F(z)$ is the averaged exchanged photon number gained from the near-field with $F(z)$ the slow-varying spatial distribution of the near field of the tip illuminated by the laser. The modulation amplitude of the n-th order multiphoton process is characterized by the Bessel function and the Gaussian envelope of momentum width $\sigma_{p_0}$, shifted relative to central momentum to $p_0 + n\delta p$.

As in the case of the unmodulated wavepacket (eq.5), we obtain the spatial evolution of the wavepacket in real space away from the tip ($z > L_c$) by substituting (10) in (2):

$$\psi(z,t) = \frac{e^{e^{i(p_0 z - \varepsilon_{p_0} t)/\hbar}}}{\left(2\pi\sigma_{z_0}^2\right)^{1/4}\sqrt{1+i\xi t}} \sum_{n=-\infty}^{\infty} J_n(2|g|) \exp\left(-\frac{(z - v_0 t - n\delta p_t/\hbar)^2}{4\sigma_{z_0}^2(1+i\xi t)}\right) e^{i\left(\frac{n\omega_b}{v_0}\right)(z - v_0 t - n\delta p_t/\hbar)} \quad (11)$$

The evolution of the modulated wavepacket along the drift section is best presented by a Wigner distribution with respect to the relative position $\zeta = z - v_0 t$ and the shifted momentum $p' = p - p_0$ that is defined as

$$W(\zeta, p', t) = \frac{1}{2\pi} \int dq \, \psi^*_{p'-q/2} \psi_{p'+q/2} \exp\left(-i\left(E_{p'+q/2} - E_{p'-q/2}\right)/\hbar\right) e^{iqz/\hbar} \quad (12)$$

Fig. 3 shows the Wigner distribution $W(\xi, p', t_D')$ after optimal (Maximum bunching) drift time $t_D' = L_D'/v_0$, where $L_D' = L_D - L_c$ is the drift length from the modulation point to the grating (see Fig. 2). We also show the projected density distributions in both momentum and spatial spaces in Fig. 4. The shown distribution parameters are $\beta_0 = 0.7$, $2|g| = 11.4$, in correspondence to Feist's experiment. [22]

The Wigner function in phase-space demonstrates the turning of momentum modulation into the tight density micro-bunching at an estimated optimal drift time [36]

$$t'_{D,\max} \approx \frac{1}{2} \frac{T_b}{\Delta p_m / p_0} \quad (13)$$

where $T_b = 2\pi/\omega_b$ is the bunching period, and the maximal effective momentum gain $\Delta p_m \approx 2|g|\delta p$ depends on the exchanged photon number at the modulation point. Note that the momentum spectrum does not vary with drift propagation, thus we cannot reveal the evolved micro-bunched structure of the modulated wavepacket by measuring its momentum spectrum alone. Finally, the probabilistic expectation of the current density of the modulated wavepacket (eq.6) after drift length $L_D'$, is found to be periodically modulated in time and space as shown in Fig. 3, which displays tight and narrow micro-bunching at the maximal bunching drift time (eq.5), with width $2\sigma_b \approx 75$ as (see in the

insert) [22]. Classically, we can write then the current density distribution in the interaction region, defined in the coordinates range $0 < z < L_G$

$$J(z,t) = |\psi(z,t)|^2 = -ef_{e\perp}(\mathbf{r}_\perp) f_e(t - z/v_0 - t_{0e}) f_{\text{mod}}(t - z/v_0 - t_0) \quad (14)$$

where $t_0$ is the modulation reference time at the entrance to the interaction region z=0. Assuming again that the wavepacket dimensions hardly change along the interaction region, and the function $f_e(t) = \dfrac{1}{(2\pi)^{1/2} \sigma_t(L_D)} e^{-t^2/2\sigma_t^2(L_D)}$ is the unmodulated wavepacket envelope (9) with $\sigma_t = \sigma_z/v_0$,

Since $f_{\text{mod}}(t)$ is periodic with period $T_b$, it can be expanded as a Fourier series in terms of the harmonics of the bunching frequency

$$f_{\text{mod}}(t) = \sum_{l=-\infty}^{\infty} B_l e^{-il\omega_b t} \quad (15)$$

where $l$ denotes the $l$-th order harmonic.

Substituting (11) in (14) we derive in Appendix A the coefficient $B_l$ of the Fourier series expansion after a drift time $t_D' = L_D'/v_0$ away from the modulation tip

$$B_l = \sum_{n=-\infty}^{\infty} J_n(2|g|) J_{n+l}(2|g|) \exp\left\{-\frac{(n\delta t_D')^2}{4\sigma_{t_0}^2(1+i\xi t_D)} - \frac{((n+l)\delta t_D')^2}{4\sigma_{t_0}^2(1-i\xi t_D)}\right\} \exp(i(2n+l)l\omega_b t_D' + il\phi_0)$$

(16)

where $\delta = \hbar\omega/m^* v_0^2$, and we kept the dependence on the initial phase $\phi_0$, which is important for the subsequent extension of the analysis to the multi-particle case, where all particle wavepackets are modulated.

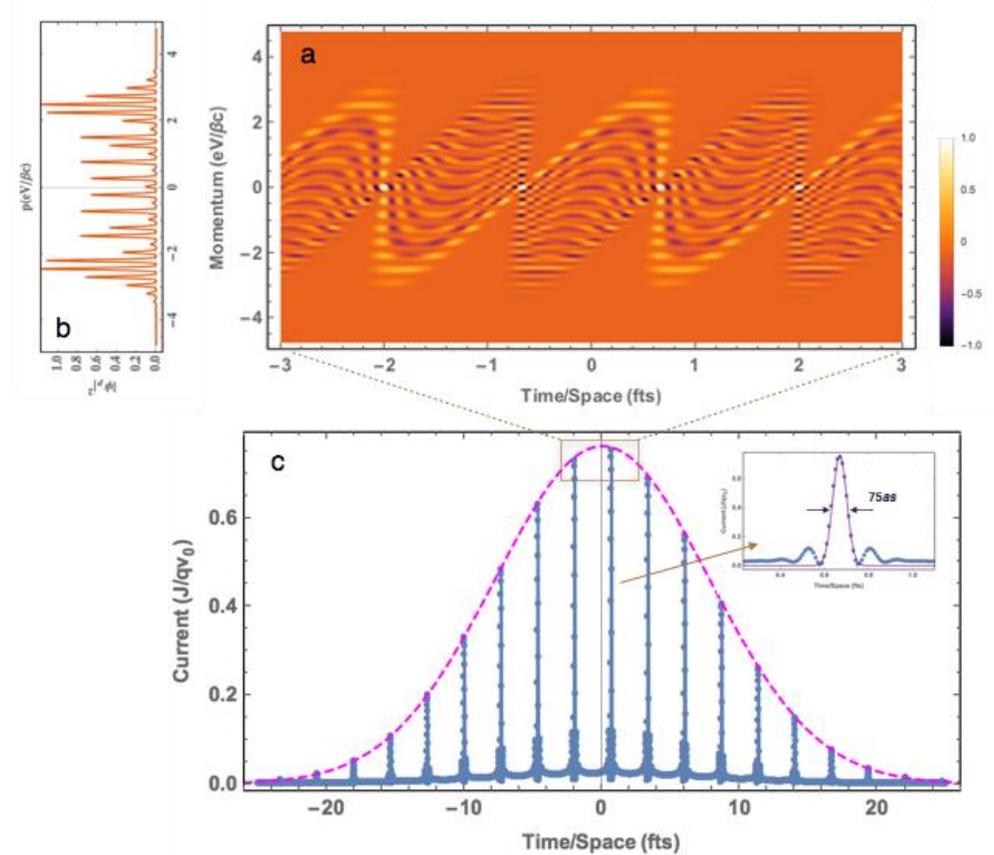

*Figure 3: Formation of attosecond tight electron density bunching. (a) The Wigner distribution after optimal drift length $L'_{D,max}$. (b) The modulated momentum (energy) distribution of electron wavepacket after interaction with the near-field on a tip. (c) The density micro-bunching of a single-electron wavepacket.*

**Formulation of the spectral optical parameters - Radiation mode expansion**

We now turn to calculate the radiation emission by the current of a beam of electron quantum wavepackets. We base our analysis on a general radiation-mode excitation formulation for bunched beam superradiance [14]. The radiation field excited by a general finite pulse of current $\mathbf{J}(\mathbf{r},\omega)$ is expanded in the frequency domain in terms of a set of orthogonal directional transverse modes $\{\tilde{\mathbf{E}}_q(\mathbf{r}), \tilde{\mathbf{H}}_q(\mathbf{r})\}$ that are the transversely confined homogeneous solution of the electromagnetic wave equations of free space or a source-less guiding structure

$$\{\mathbf{E}(\mathbf{r},\omega),\mathbf{H}(\mathbf{r},\omega)\} = \sum_q C_q(\omega,z)\{\tilde{\mathbf{E}}_q(\mathbf{r}),\tilde{\mathbf{H}}_q(\mathbf{r})\} \qquad (17)$$

where $C_q(\omega,z)$, the slowly growing field amplitude along the propagation direction (z) of a radiation mode q at spectral frequency $\omega$ is derived from Maxwell equations [14]. The increment of the field amplitude of mode q is

$$\Delta C_q(\omega) = C_q^{out}(\omega) - C_q^{in}(\omega) = -\frac{1}{4P_q}\int \mathbf{J}(\mathbf{r},\omega)\cdot\tilde{\mathbf{E}}_q^*(\mathbf{r})d^3\mathbf{r} \qquad (18)$$

where $P_q = \frac{1}{2}\Re\iint \tilde{\mathbf{E}}_q(\mathbf{r})\times\tilde{\mathbf{H}}_q(\mathbf{r})\cdot\hat{e}_z d^2r_\perp$ is the normalization power of mode q.

The spectral radiative energy emission per mode, derived from Wiener-Khincine theorem (see Appendix B), is given (for $\omega>0$) by

$$\frac{dW_q(\omega)}{d\omega} = \frac{2}{\pi}P_q\left|C_q^{out}(\omega)\right|^2 \qquad (19)$$

Substituting (18) in (19), the emitted spectral radiative energy per mode can be written in terms of three parts

$$\frac{dW_q(\omega)}{d\omega} = \left(\frac{dW_q(\omega)}{d\omega}\right)_{in} + \left(\frac{dW_q(\omega)}{d\omega}\right)_{SP/SR} + \left(\frac{dW_q(\omega)}{d\omega}\right)_{ST-SR} \qquad (20)$$

with

$$\left(\frac{dW_q(\omega)}{d\omega}\right)_{in} = \frac{2}{\pi}P_q\left|C_q^{in}(\omega)\right|^2$$

$$\left(\frac{dW_q(\omega)}{d\omega}\right)_{SP/SR} = \frac{2}{\pi}P_q\left|\Delta C_q(\omega)\right|^2 = \left|\int \mathbf{J}(\mathbf{r},\omega)\cdot\tilde{\mathbf{E}}_q^*(\mathbf{r})d^3\mathbf{r}\right|^2 \qquad (21)$$

$$\left(\frac{dW_q(\omega)}{d\omega}\right)_{ST-SR} = \frac{4}{\pi}P_q\Re\left\{C_q^{in*}(\omega)\Delta C_q(\omega)\right\} = -\frac{1}{\pi}\Re\left\{C_q^{in*}(\omega)\int \mathbf{J}(\mathbf{r},\omega)\cdot\tilde{\mathbf{E}}_q^*(\mathbf{r})d^3\mathbf{r}\right\}$$

where the first term is the spectral energy of the input radiation wave; the second term corresponds to radiation emission independent of the input wave – random spontaneous (SP) or coherent (superradiant) (SR); the third term is stimulated-superradiance (ST-SR) corresponding to stimulated interaction (emission/absorption) of the input radiation wave $C_q^{in}(\omega)$ and the spectral component of the current. A detailed derivation of the radiation mode expansion formulation is given in Appendix B.

**Results and Discussion**

### I. Radiation of a Single electron wavepacket

### A. Unmodulated quantum wavepacket

First we consider the simple case of the probabilistic current of a single electron-wavepacket (eq. 6) in the Fourier transform frequency domain

$$\mathbf{J}(\mathbf{r},\omega) = \int_{-\infty}^{\infty} dt\, \mathbf{J}(\mathbf{r},t) e^{i\omega t} \tag{22}$$

We use eqs. (9) for the current of the unmodulated electron wavepacket, and then, after Fourier transformation (using $F\{e^{-t^2/2\sigma_t^2}/\sqrt{2\pi}\} = e^{-\omega^2\sigma_t^2/2}$), substitute it into (18)

$$\Delta C_q(\omega) = \frac{e}{4P_q}\tilde{M}_q(\omega) e^{-\sigma_t^2\omega^2/2} e^{i\omega t_{0e}} \tag{23}$$

where we defined here an overlap integral parameter (analogous to "matrix element" in spatial space)

$$\tilde{M}_q(\omega) = \int f_e(\mathbf{r}_\perp)\tilde{E}_{qz}^*(\mathbf{r}_\perp) e^{i\omega(z/v)} d^3r \tag{24}$$

Substituting (23) in (21), we get the expressions for the spontaneous and stimulated emission of a single wavepacket

$$\left(\frac{dW_q}{d\omega}\right)_{e,SP} = \frac{e^2}{8\pi P_q}\left|\tilde{M}_q(\omega)\right|^2 e^{-\omega^2\sigma_t^2}$$

$$\left(\frac{dW_q}{d\omega}\right)_{e,ST} = \frac{e}{\pi}\mathrm{Re}\left\{C_q^{in*}(\omega)\tilde{M}_q(\omega)e^{i\omega t_{0e}}\right\}e^{-\omega^2\sigma_t^2/2} \tag{25}$$

and we see right away that both spontaneous and stimulate emission expressions are wavepacket-size-dependent in the semi-classical regime, and vanish in the quantum wavepacket limit $\omega\sigma_t \gg 1$ (Eq. 1). We further reduce these expressions in the case where the axial component of the radiation mode is a traveling wave of wavenumber $q_z$

$$\tilde{E}_{qz}(z,\mathbf{r}_\perp) = \tilde{E}_{q_z}(\mathbf{r}_\perp)e^{iq_z z} \tag{26}$$

Then the axial integration in (eq.24) can be carried out

$$\tilde{M}_q = \tilde{M}_{q\perp} E_{qz0} L e^{i\theta L/2} \mathrm{sinc}(\theta L/2) \tag{27}$$

where we defined a normalized coefficient describing the transverse overlap between the field of the radiation mode and the electron wavepacket

$$\tilde{M}_{q\perp} = \frac{1}{E_{qz0}} \int f_e(\mathbf{r}_\perp) \tilde{E}_{q_z}^*(\mathbf{r}_\perp) d^2 r_\perp$$

Here $E_{qz0} = \left|\tilde{E}_{qz}(\mathbf{r}_{\perp 0})\right|$, where $\mathbf{r}_{\perp 0}$ is the transverse coordinate of the center of the wavepacket profile.

The parameter

$$\theta = \frac{\omega}{v_0} - q_z(\omega) \tag{28}$$

is the electron/radiation-wave synchronism (detuning) parameter. In these terms, the spontaneous emission and stimulated emission are given by

$$\left(\frac{dW_q}{d\omega}\right)_{e,SP} = \frac{e^2 E_{qz0}^2 L^2}{8\pi P_q} \left|\tilde{M}_{q\perp}\right|^2 \mathrm{sinc}^2(\theta L/2) e^{-\omega^2 \sigma_t^2} = W_q \mathrm{sinc}^2(\theta L/2) e^{-\omega^2 \sigma_t^2}$$

$$\left(\frac{dW_q}{d\omega}\right)_{e,ST} = \frac{e}{\pi} E_{qz0} L \, \mathrm{Re}\left\{C_q^{in*}(\omega)\tilde{M}_{q\perp} e^{i\omega t_{0_e}+i\theta L/2}\right\} \mathrm{sinc}(\theta L/2) e^{-\omega^2 \sigma_t^2/2} \tag{29}$$

$$W_q = \frac{e^2 E_{qz0}^2 L^2}{8\pi P_q} \left|\tilde{M}_{q\perp}\right|^2 \qquad 29A$$

Note that if the transverse wavepacket function is narrow relative to the transverse variation of the field, then $M_{q_\perp} = 1$. Also note that for a long interaction length L, efficient interaction can take place only if (eq.26) is a slow-wave radiation field component (e.g. in Cerenkov radiation or Smith-Purcell radiation), where a synchronism condition $v_{ph} = \omega/q_z \approx v$ can be established, so that $\theta \simeq 0$. Wide frequency band emission can take place also in a short interaction length $\theta L \ll 1$ without satisfying a synchronizm condition (e.g. in transition radiation). In the limit $\omega \sigma_t \ll 1$, Eq. 29 reduces to the classical expression for spontaneous emission of a single point particle [14].

## B. "Einstein Relations" and spectral correspondence of energy exchange conservation in electron interaction with radiation

Now let us concentrate on the stimulated emission term. Assume that the interacting field component of the incident wave is a single frequency harmonic wave (e.g. a laser beam field)

$$E_{int} = E_0 \cos(\omega_0 t + \varphi_0) \tag{30}$$

In terms of the continuous spectral formulation (eq. 17) and spectral normalization of (eq.18) for $\omega > 0$, this corresponds to (see Appendix C):

$$C_q^{in}(\omega) = \pi E_0 e^{-i\varphi_0} \delta(\omega - \omega_0) \tag{31}$$

Then from integration of eq. 21 over $\omega$, the incremental stimulated-emission radiation energy from a single electron wavepacket is

$$(\Delta W_q)_{e,ST} = eE_0 L |\tilde{M}_{q_\perp}| \cos(\theta_0 L/2 + \omega_0 t_{0e} - \varphi_0) \operatorname{sinc}(\theta_0 L/2) e^{-\omega_0^2 \sigma_t^2/2} \tag{32}$$

where $\theta_0 = \theta(\omega_0)$. This radiative energy gain/loss is in complete agreement with the energy loss/gain of a single electron quantum wavepacket as calculated semi-classically by the solution of Schrodinger equation in [27]. It is also consistent with the classical point-particle limit [14] when $\sigma_t \ll 1/\omega_0$. This shows that conservation of energy exchange between a coherent radiation field (laser) and an electron wavepacket, contained and interacting entirely within the spatial volume of a single radiation mode, is maintained within the minimal spectral phase-space volume representing the coherent single radiation mode:

$$\left(\Delta W_q\right)^{RAD}_{e,ST} = -\left(\Delta W_q\right)^{GAIN}_{e,ST}$$

Another important new result is a universal relation between stimulated emission radiative energy gain at frequency $\omega_0$ and spontaneous emission spectral radiant energy into the same coherent phase space volume (single radiation mode) at the same frequency. At maximum emission (synchronous) interaction condition $\theta = 0$ and maximum deceleration phase of the electron wavepacket relative to the wave $\omega_0 t_{0e} - \varphi_0 = 0$, this relation is

$$\left(\Delta W_q(\omega_0)\right)^2_{e,ST,max} = \frac{8\pi E_0^2}{E_{qz0}^2 / P_q} \left(\frac{dW_q(\omega_0)}{d\omega}\right)_{SP,max} \quad (33)$$

This universal relation is only valid in the classical point-particle limit and in the quantum to classical transition range of the wavepacket $\sigma_t \lessapprox 1/\omega$. In the opposite, inherent quantum wavepacket limit, $\omega \sigma_t \gg 1$ (eq.1), both stimulated and spontaneous emission expressions vanish. Note that this semi-classical "Einstein relation" between classical spontaneous emission and stimulated emission is different from the classical limit relation between stimulated emission and quantum spontaneous emission derive in [15] in a QED model. Of course, the semi-classical analysis of an electron wavepacket cannot produce the quantum spontaneous emission. This aspect is addressed in a companion article based on QED formulation [28].

It is instructive to observe that the proportionality coefficient in eq. 33 can be related to Pierce's known "interaction impedance" parameter $K_{Pierce} = E_{qz0}^2 / 2q_z P_q$ [29]. Note that in order to use the relation (33) in practice, e.g. for the case of Smith-Purcell radiation, one must solve first analytically or numerically the classical electromagnetic problem of the amplitude of the interacting axial field component $E_{qz}$ relative to the normalization power of the entire mode - $P_q$ [37].

### C. Modulated quantum wavepacket

Secondly, we consider the case of a modulated electron wavepacket. In the case of a modulated wavepacket, using (eqs.14-15,22) one gets

$$\Delta C_q(\omega) = \frac{e}{4P_q} e^{i\omega z/v_0} \tilde{M}_q \cdot \sum_{l=-\infty}^{\infty} B_l e^{-\sigma_t^2(\omega-l\omega_b)^2/2} e^{i(\omega-l\omega_b)t_{0e}} e^{il\omega_b t_0} \equiv \frac{e\tilde{M}_q B(\omega)}{4P_q} \quad (34)$$

where $B(\omega) = e^{i\omega z/v_0} \sum_{l=-\infty}^{\infty} B_l e^{-\sigma_t^2(\omega-l\omega_b)^2/2} e^{i(\omega-l\omega_b)t_{0e}} e^{il\omega_b t_0}$. Figure 4 shows the harmonics of the current bunching amplitude factor $B(\omega)$ as function of frequency (a) and drift time $t_D'$ (b) for optically-modulated electron wavepacket, where $B_l$ was evaluated from eq. 16.

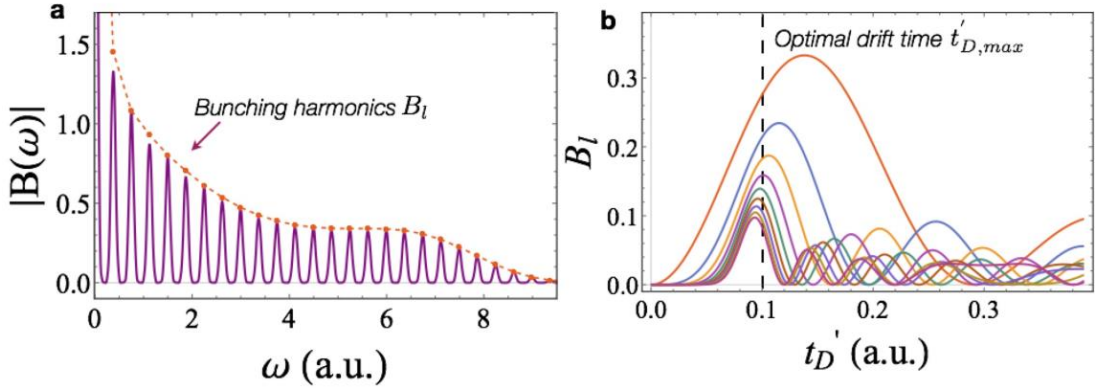

Figure 4: (a) The current bunching factor $B(\omega)$ as function of frequency for optically-modulated electron wavepacket at the optimal drift time $t_{D,max}'$. (b) The dependence of the $l^{th}$-order harmonic bunching factor $B_l$ on the drift time $t_D'$. The optimal drift time $t_{D,max}'$ is marked by the vertical dashed-line.

Using eq. 35 in eq. 21, the expressions for spontaneous emission by a single electron modulated wavepacket is:

$$\left(\frac{dW_q(\omega)}{d\omega}\right)_{e,SP-MOD} = \sum_l \left(\frac{dW_{q,l}(\omega)}{d\omega}\right)_{e,SP-MOD} \qquad (35)$$

where

$$\left(\frac{dW_{q,l}(\omega)}{d\omega}\right)_{e,SP-MOD} = W_q |B_l|^2 \operatorname{sinc}^2(\theta L/2) e^{-\sigma_t^2(\omega - l\omega_b)^2} \qquad (36)$$

where in (35) we assumed that the overlaps between the spectral lines of the harmonics $l$ are negligible, as shown in Figure 4.a.

The incremental stimulated energy emission/absorption of a modulated electron wavepacket in the case of a coherent incident radiation field (eq.30-31) is

$$\left(\Delta W_q\right)_{e,ST-MOD} = \sum_l \left(\Delta W_{q,l}\right)_{e,ST-MOD}$$

where

$$\left(\Delta W_{q,l}\right)_{e,ST-MOD} = eE_0 L |\tilde{M}_{q\perp}||B_l| \cos\left[\theta_0 L/2 + (\omega_0 - l\omega_b)t_{0e} + l\omega_b t_0 - \varphi_0\right] \operatorname{sinc}(\theta_0 L/2) e^{-\sigma_t^2(\omega_0 - l\omega_b)^2/2}$$
(37)

where $\theta_0 = \theta(\omega_0) = \omega_0/v - q_z(\omega_0)$. A striking difference between these expressions and the corresponding cases for the unmodulated quantum wavepacket (29, 32) is that the modulated wavepacket can emit/absorb radiation at frequencies beyond the quantum cut-off condition (1), which occur at all harmonic frequencies $l\omega_b$ of the wavepacket modulation of significant component amplitude $B_l$.

### D. Smith-Purcell Radiation (SPR) - An Example

A vivid presentation of radiation emission extinction and revival effects of a modulated quantum electron wavepacket is presented here for the case of the Smith-Purcell radiation experiment as shown in Figure 1&2. The modes of the SPR grating structure are Floquet modes

$$\tilde{E}_q(\mathbf{r}) = \sum_m \tilde{E}_{qm}(\mathbf{r}_\perp) e^{i(q_{z0}+mk_G)z} \qquad (38)$$

where $k_G = 2\pi/\lambda_G$, $\lambda_G$ is the grating period and

$$q_{z0} = \left(\omega^2/c^2 - k_{\perp q}^2\right)^{1/2} = \frac{\omega}{c}\cos\Theta_q \qquad (39)$$

The angle $\Theta_q$ is the 'zig-zag' angle of mode q in a waveguide structure. We use in (eq. 26) $q_z = q_{z0} + mk_G$ where m is the m-th order space harmonic of the Floquet mode, and apply all the expressions for spontaneous and stimulated emission to each of the space harmonics with a detuning parameter (neglecting the interference between the space harmonics)

$$\theta_m = \frac{\omega}{v} - q_{z_0} - mk_G \qquad (40)$$

The spontaneous emission expression (eq. 29) for an unmodulated wavepacket and Eqs. 36-37 for a modulated wavepacket, can be modified to include interaction with any space-harmonics m, which can be synchronous with the electron $\omega/(q_{z0}+mk_G) \approx v$

$$\begin{aligned}\left(\frac{dW_q(\omega)}{d\omega}\right)_{e,SP} &= \sum_m \left(\frac{dW_{q,m}(\omega)}{d\omega}\right)_{e,SP} = \sum_m W_{q,m}\mathrm{sinc}^2(\theta_m L/2) e^{-\omega^2 \sigma_t^2} \\ \left(\frac{dW_q(\omega)}{d\omega}\right)_{e,SP-MOD} &= \sum_{l,m} \left(\frac{dW_{q,lm}(\omega)}{d\omega}\right)_{e,SP-MOD} = \sum_{l,m} W_{q,m}|B_l|^2 \mathrm{sinc}^2(\theta_m L/2) e^{-\sigma_t^2(\omega-l\omega_b)^2}\end{aligned} \qquad (41)$$

$$W_{q,m} = \frac{e^2 E_{qz,m}^2 L_G^2}{8\pi P_q} \left|\tilde{M}_{q_\perp,m}\right|^2$$

Figure 5a displays the SPR spectrum in terms of wavelength $\lambda = 2\pi c/\omega$ and angle $\Theta$ in the classical limit $\sigma_t \ll 1/\omega$, where the wavepacket appears as a point-particle. Emission lines appear for arbitrary angles in the range $0 < \Theta < \pi$ at frequencies or wavelengths corresponding to the synchronism condition $\theta(\omega_m) = 0$

$$\omega_m(\Theta) = mck_G/(\beta^{-1} - \cos\Theta), \quad \text{or} \quad \lambda_m(\Theta) = \frac{\lambda_G}{m}(\beta^{-1} - \cos\Theta) \tag{42}$$

in agreement with the classical SPR emission relation. [6]

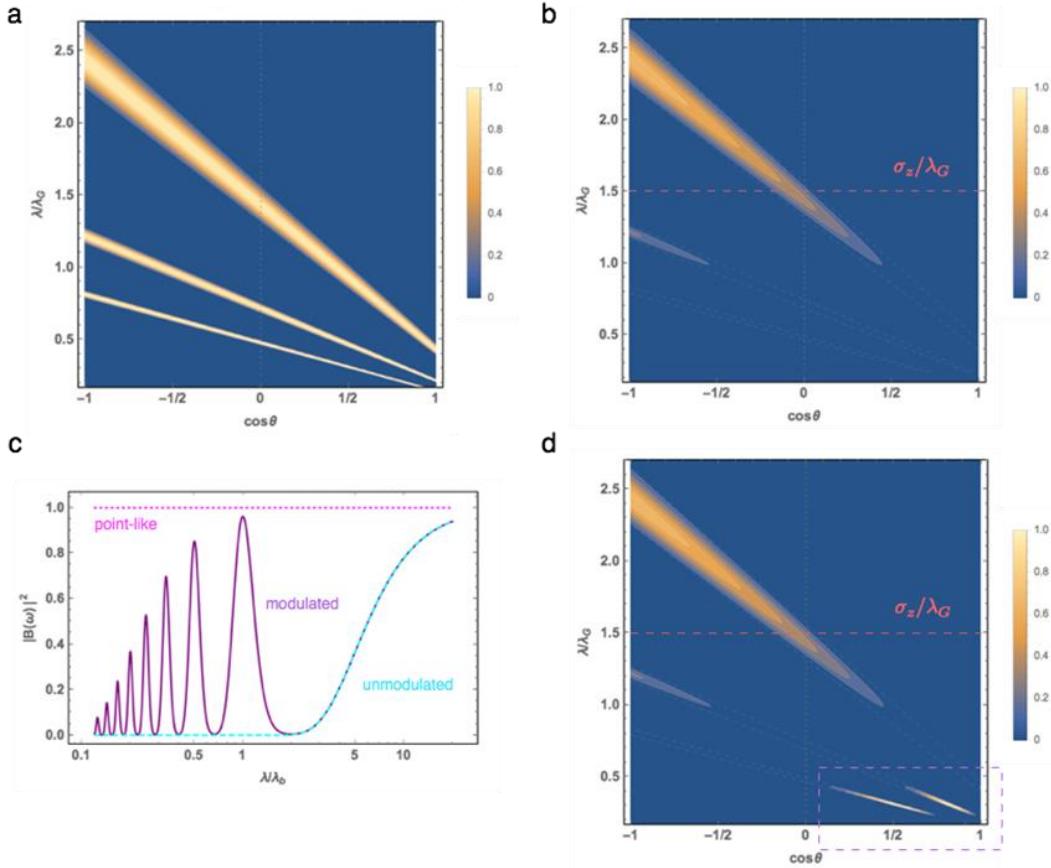

*Figure 5: The wavepacket limit of Smith-Purcell spectrum. a The classical SP spectrum for point-like particles. b The effect of short wavelength cut-off of a Gaussian wavepacket. d The same for a modulated Gaussian wavepacket, displaying appearance of non-classical superradiant-SP harmonic radiation light-spots at the short wavelength part. c The modeled bunching factors of point-like, un-modulated and modulated currents of electron wavepacket as a function of relative wavelength $\lambda/\lambda_G$.*

In Fig.5b, we show the expected SPR spontaneous emission spectrum of an unmodulated electron wavepacket in a set-up shown in Fig. 1. The plot shows that at low frequencies $\omega < 1/\sigma_t$, or long wavelengths $\lambda > 2\pi c\sigma_t$, the low-pass filtering effect of the finite size wavepacket extinction parameter $e^{-\omega^2\sigma_t^2}$ of eq.41A (Fig. 5c, blue curve) keeps the long wavelength part of the classical SPR spectrum of fig.5a unaffected, and cuts off the shorter wavelength and higher harmonic sections when $1/\omega_{m-1}(\Theta=0)<\sigma_t<1/\omega_{m-1}(\Theta=\pi)$, namely,

$$(1-\beta_0)\frac{\lambda_G}{2\pi}<\sigma_z<(1+\beta_0)\frac{\lambda_G}{2\pi} \tag{43}$$

This case is displayed in Fig. 5b&c for the parameters $\sigma_z/\lambda_G=0.22, \beta_0=0.7, N_G=L_G/\lambda_G=9$. Here the first order SPR harmonic is partly cut-off, the second order harmonic is barely observable and higher harmonics are extinct.

However, a more dramatic change in the spectrum takes place when the wavepacket is modulated (the "modulating laser" in Fig.2 is turned "ON"). In this case the wavepacket bunching factor $e^{-\sigma_t^2(\omega-l\omega_b)^2}$ in eq. 41 permits resonant emission only at harmonics $\omega_l = l\omega_b$ and this harmonics-spectrum is cut-off only at much higher frequencies by the filtering effect of the narrow micro-bunches $l\omega_b<1/\sigma_t$, or

$$l<\frac{1}{2\pi}\frac{T_b}{\sigma_t} \tag{44}$$

Figure 5c displays the (classical) SPR spontaneous emission spectrum (eq.43) in terms of $\lambda$, for the same parameters and wavepacket size $\frac{\lambda}{\lambda_G}=1.5$ as in the un-modulated wavepacket case (Fig.5b). It is seen that the bunching factor exhibits resonance at harmonics of the bunching frequencies $\omega_b$ in the frequency range that was cut-off before modulation. This shows up in Fig.6 as three resonant spots, reviving the 1$^{st}$, 2$^{nd}$ and 3$^{rd}$ order SP space harmonics at distinct emission angles.

Inspection of Eq.43 reveals that resonant emission spots will appear, within the spectral range above the frequency cut-off of the unmodulated beam radiation, $\omega>1/\sigma_t (\lambda<2\pi c\sigma_t)$ at the frequencies and emission angles in which the narrow band filtering function $e^{-\sigma_t^2(\omega-l\omega_b)^2}$ and $\text{sinc}^2(\theta_m(\omega)L_G/2)$ overlap. The centers of these spots are at positions $\omega=\omega_l, \Theta=\Theta_{lm}$, where $\Theta_{lm}$ is the solution of the equation

$$l\omega_b = \omega_m(\Theta_{lm}) \qquad (45)$$

and $\omega_m(\Theta)$ is given by Eq. 42. The spectral width of the spot depends on which filtering function is narrower. The spectral width of the synchronization function is

$$\left.\frac{\Delta\omega}{\omega_m}\right|_\Theta = \frac{1}{mN_G} \qquad (46)$$

and of the bunched wavepacket bunching factor

$$\frac{\Delta\omega}{\omega_l} = \frac{2}{\pi}\frac{1}{lN_b} \qquad (47)$$

where $N_b = 2\sigma_t/T_b$ is the number of micro-bunches in the modulated wavepacket.

Figure 6 displays the emission spectrum for the two opposite cases $N_G > N_b$ and $N_G < N_b$ as functions of $\lambda$ and $\Theta$.

The same configuration of Smith-Purcell experiment can be used for measurement of stimulated emission/absorption and corresponding deceleration/acceleration with the interaction "input laser" beam turned "ON". In this case, the incremental exchanged energy from the electron wavepacket to the radiation wave and vice versa is given by a modified version of Eqs. 32&37 (for the unmodulated and modulated cases respectively)

$$(\Delta W_q)_{e,ST} = \sum_m (\Delta W_q)_{e,ST,m} = eE_0 L_G \sum_m |\tilde{M}_{q_\perp,m}| \cos(\theta_m L_G/2 + \omega_0 t_0 - \varphi_0)\text{sinc}(\theta_m L_G/2) e^{-\sigma_t^2\omega^2/2} \qquad (48)$$

$$\left(\Delta W_q\right)_{e,ST-MOD} = \sum_{m,l}\left(\Delta W_{q,l,m}\right)_{e,ST-MOD}$$

$$= eE_0 L_G \sum_{m,l}\left|\tilde{M}_{q_\perp,m}\right| B_l \cos\left[\theta_m(\omega_0)L/2 - (\omega_0 - l\omega_b)t_{0e} + l\omega_b t_0 - \varphi_0\right]\operatorname{sinc}\left[\theta_m(\omega_0)L_G/2\right] e^{-\sigma_t^2(\omega_0 - l\omega_b)^2/2}$$

(49)

Evidently a spectral diagram similar to Fig.5-6, with cutoff effects and re-emerging spots (in case of a modulated beam), would be measured in the incremental energies of the radiation wave and the electrons when the incident laser beam is scanned over wavelengths $\lambda$ and incident angle $\Theta$.

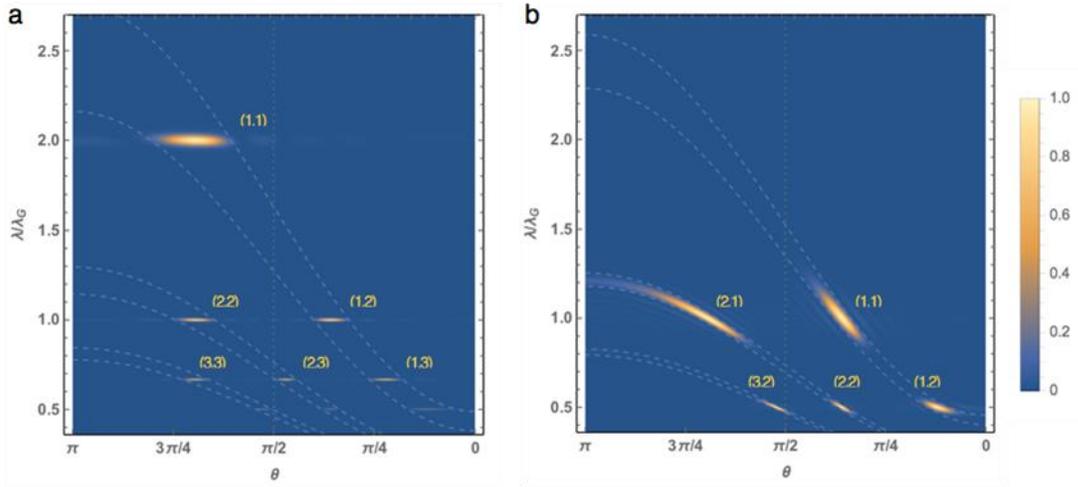

*Figure 6: Beyond-cutoff Smith-Purcell radiation (SPR) spectrum of a modulated quantum electron wavepacket. (a) SPR for $N_G > N_b$. (b) SPR for $N_G < N_b$.*

It is argued that this scheme and the characteristic spectral map can be used for measuring the electron quantum wavepacket size $\sigma_t$ at the entrance to the grating. We note, however, that the semi-classical calculation of spontaneous emission from a single electron quantum wavepacket may have limited validity, as discussed in the companion paper [28] based on QED formulation, and its measurement may be jeopardized by quantum spontaneous emission noise, not inclusive in a semi-classical formulation. On the other hand, the validity of using semi-classical formulation for stimulated radiative interaction is well founded. We therefore assert that such a <u>stimulated interaction</u> experiment can be a way for measuring the quantum wavepacket size with a SP experiment as shown in Figure 1-2. The available control over the input radiation field intensity $E_0$ can help to overcome expected quantum and other noise factors in the experimental measurement.

## II. Radiation of multi-electron wavepacket beams

Beyond the single electron radiation cases, we now consider the radiation of a multi-particle wavepackets beam ( $N \gg 1$ ). Radiation measurements with single electron wavepacket would be challenging experiments. To get significant wavepacket-dependent measurement, repeated experiments must be performed with careful pre-selection filtering, to assure similarity (or identity) of the wavepackets in successive measurement experiments [38]. We now consider the case where we measure at once a pulse of electron wavepackets that may be correlated at entrance to the radiative interaction region (see Fig. 7). Now, assume that the e-beam is composed of electron-wavepackets whose current density distribution is given by $\mathbf{J}(\mathbf{r},\omega) = \sum_{j=1}^{N} \mathbf{J}_j(\mathbf{r},\omega)$. Consequently, the spontaneous/superradiant emission and stimulated-superradiant emission of the beam (eq. 21) are respectively

$$\left(\frac{dW_q}{d\omega}\right)_{SP/SR} = \frac{2}{\pi} P_q \left| \sum_{j=1}^{N} \Delta C_{qj}(\omega) \right|^2$$
$$\left(\frac{dW_q}{d\omega}\right)_{ST-SR} = \frac{4}{\pi} P_q \operatorname{Re}\left\{ C_q^{in*}(\omega) \sum_{j=1}^{N} \Delta C_{qj}(\omega) \right\}$$
(50)

where we define as in (18)

$$\Delta C_{qj}(\omega) = -\frac{1}{4P_q} \int \mathbf{J}_j(\mathbf{r},\omega) \cdot \tilde{\mathbf{E}}_q^*(\mathbf{r}) d^3r \tag{51}$$

and the current $\mathbf{J}_j(\mathbf{r},\omega)$ of the j electron-wavepacket is given by

$$J_j(r,\omega) = \int_{-\infty}^{\infty} dt\, J_j(r,t) e^{i\omega t}$$

### E. Quantum electron wavepackets beam

We go back to eqs.21-25 and consider the cooperative emission of a pulse of N electrons. Assuming all wavepackets are identical (except for arrival time $t_{0j}$), the averaged spectral energy of the pulse is

$$\left\langle \left(\frac{dW_q}{d\omega}\right)^{pulse}_{SP/SR} \right\rangle \equiv \left\langle \left|\sum_{j=1}^{N} e^{i\omega t_{0j}}\right|^2 \right\rangle \left(\frac{dW_q}{d\omega}\right)_{e,SP}$$
$$\left\langle \left(\frac{dW_q}{d\omega}\right)^{pulse}_{ST-SR} \right\rangle \equiv \left\langle \sum_{j=1}^{N} e^{i\omega t_{0j}} \right\rangle \left(\frac{dW_q}{d\omega}\right)_{e,ST} \quad (52)$$

where the single electron spontaneous and stimulated spectral energies are given in (29). It is evident that when the electron beam pulse is longer than the radiation optical period $\sigma_{pulse} > T = 2\pi/\omega$, and $t_{oj}$ are random, all the phasor terms in (52B) and all the mixed terms in (52A) interfere destructively. One gets then no average stimulated-emission (and no average acceleration) of the random electron beam, but there is a resultant "classical spontaneous emission" (Shot-noise radiation) of the beam, originating from the diagonal terms in the product in Eq. 52A. Using (29A)

$$\left\langle \left(\frac{dW_q}{d\omega}\right)^{pulse}_{SP} \right\rangle = N \left(\frac{dW_q}{d\omega}\right)_{e,SP} = NW_q \text{sinc}^2(\theta L/2) e^{-\omega^2 \sigma_t^2} \quad (53)$$

Namely, the spontaneous emission from N particle is N times the spectral emission from a single electron. In consideration of the high frequency cut-off of the single electron spontaneous emission Eq. 1, one concludes that spontaneous "shot-noise" radiation of an electron beam is not "white noise", but diminishes in the quantum limit $\omega \sigma_t \gg 1$ [30].

In the opposite limit of a short electron beam pulse relative to the optical period $\sigma_{pulse} < T = 2\pi/\omega$, all phasor terms in (52) within the pulse, sum-up with the same phase and since necessarily $\sigma_t < \sigma_{pulse}$, the exponential decay factors in (29) are unity, and the electrons radiate as point particles without any dependence on $\sigma_t$. This is actually the classical case of superradiance analyzed in [14], where the collective emission of the electron pulse depends only on the particles arrival time distribution in the coefficients of eq. 52,

$$b_{pulse} = \frac{1}{N}\left\langle \sum_{j=1}^{N} e^{i\omega t_{0j}} \right\rangle \tag{53A}$$

One can replace the summation over j by integration over the temporal distribution of the particles in the beam pulse $f_{pulse}(t - t_{0,pulse})$, where $\int_{-\infty}^{\infty} f_{pulse}(t_{0j}) dt_{0j} = 1$ [30,14] (see Append. D). For a Gaussian pulse temporal distribution

$$b_{pulse} = \int_{-\infty}^{\infty} f_{pulse}(t_{0j} - t_{0,pulse}) e^{i\omega t_{0j}} dt_{0j} = e^{-\sigma_{pulse}^2 \omega^2 / 2} e^{i\omega t_{0,pulse}} \tag{54}$$

and

$$\left| b_{pulse} \right|^2 \approx e^{-\sigma_{pulse}^2 \omega^2} \tag{55}$$

Using these expressions and Eq. 29 with $\sigma_t = 0$ in (52)

$$\left\langle \left(\frac{dW_q}{d\omega}\right)_{SR}^{pulse} \right\rangle = N^2 \left(\frac{dW_q}{d\omega}\right)_{e,SP} = N^2 W_q \text{sinc}^2(\theta L/2) e^{-\sigma_{pulse}^2 \omega^2}$$

$$\left\langle \left(\frac{dW_q}{d\omega}\right)_{ST-SR}^{pulse} \right\rangle = N \left(\frac{dW_q}{d\omega}\right)_{e,ST} = NeE_{qz0} L \text{Re}\left\{ C_q^{in*}(\omega) \tilde{M}_{q\perp} e^{i\omega t + i\theta L/2} \right\} \text{sinc}(\theta L/2) e^{-\sigma_{pulse}^2 \omega^2 / 2} \tag{56}$$

Which is the same as the single electron expressions (29) with respective $N^2$ and $N$ factors, and $\sigma_t$ replaced by $\sigma_{pulse}$. Thus, we obtained in this limit the super-radiance and stimulated-super-radiance expressions of classical point-particles $\omega < 1/\sigma_{pulse}$ [14]. The classical point particle limit of stimulated superradiance (eq. 56B) is nothing but the acceleration formula in conventional particle accelerators. The classical superradiance formula (eq. 56A) is of interest primarily for THz radiation generation devices, since attainable short electron beam pulse durations ($t_{pulse}$) are in the order of $10^{-12}$s [39]. It is of less interest in the present context, since it has no quantum wavepacket dependence.

## F. Modulated-wavepackets electron beam: Superradiance

Consider a case where all electron wavepackets in an electron beam pulse are modulated in phase correlation by the same coherent laser beam. In this case, the modulated currents of the wavepackets in the pulse (eq. 14) are also correlated, and their radiation emissions in the interaction region are phase correlated as well (see Fig. 7).

Assume that the expectation value of the electron wavepackets probability density of an ensemble of N electrons is modulated coherently at frequency $\omega_b$ by a laser beam, as shown in Fig. 2; the single electron wavepacket density function in (14) is for electron e=j:

$$J_j(z,t) = -e f_{e\perp}(\mathbf{r}_\perp) f_{en}(t - z/v_0 - t_{0j}) f_{mod}(t - z/v_0 - t_0) \tag{57}$$

where $f_{mod}(t)$ is the periodic modulation function (15), composed of many harmonics of $\omega_b$. Important to note that the modulation phase $\omega_b t_0$ is <u>common to all the wavepackets</u> (determined by the modulating laser phase). Here $f_{en}(t)$ is the wavepacket density envelope given by the Gaussian (eq.8). The incremental spectral amplitude $\Delta C_q(\omega)$ due to interaction with the entire e-beam pulse is then found by setting e=j, and summing up $\Delta C_{qj}(\omega)$ (Eq. 34) over all particles

$$\Delta C_q(\omega) = \sum_{j=1}^{N} \Delta C_{q_j}(\omega) = -\frac{e}{4P_q} E_{q_z 0} \tilde{M}_q(\omega) \cdot \sum_l B_l e^{-\sigma_t^2(\omega - l\omega_b)^2/2} e^{il\omega_b t_0} \sum_{j=1}^{N} e^{i(\omega - l\omega_b)t_{0j}} \tag{58}$$

We now define the beam bunching factor of $l^{th}$-order harmonic frequency over the entire pulse

$$b_{pulse}^{(l)} = \frac{1}{N} \sum_{j=1}^{N} e^{i(\omega - l\omega_b)t_{0j}} \tag{59}$$

and for $N \gg 1$ replace the summation with integration, over the pulse temporal distribution function. For a Gaussian distribution $f_{pulse}(t_{0j}) = \frac{1}{(2\pi)^{1/2} \sigma_{pulse}} e^{-t_{0j}^2/2\sigma_{pulse}^2}$ (see Append. D):

$$\left\langle b_{pulse}^{(l)} \right\rangle = e^{-\sigma_{pulse}^2 (\omega - l\omega_b)^2 / 2} \tag{60}$$

$$\left\langle \left| b_{pulse}^{(l)} \right|^2 \right\rangle \approx e^{-\sigma_{pulse}^2 (\omega - l\omega_b)^2} \tag{61}$$

Substitution of eqs. 27,34 in (50) results in

$$\left( \frac{dW_q(\omega)}{d\omega} \right)_{SR-MOD}^{pulse} = \sum_l \left( \frac{dW_{q,l}(\omega)}{d\omega} \right)_{SR-MOD}^{pulse} \tag{62}$$

$$\left( \frac{dW_{q,l}(\omega)}{d\omega} \right)_{SR-MOD}^{pulse} \simeq N^2 W_q |B_l|^2 \operatorname{sinc}^2(\theta L/2) e^{-\sigma_{pulse}^2 (\omega - l\omega_b)^2} \tag{63}$$

for all harmonics $l \neq 0$. The term $l = 0$ that corresponds to spontaneous emission from a random unmodulated beam, but with a factor $|B_0|^2$, and just as in eq. 53, it decays as $e^{-\sigma_t^2 \omega^2}$ and cuts-off for $\omega > 1/\sigma_t$. All other harmonics can radiate beyond this cutoff (as in Fig 5 in the Smith-Purcell example) with a narrow bandwidth filtering-factor $e^{-\sigma_{pulse}^2(\omega-l\omega_b)^2}$ similar to the single electron modulated wavepacket case (Eq. 36), but with the electron-beam pulse duration $\sigma_{pulse}$ replacing the quantum wavepacket size parameter $\sigma_t$. Note that even though the modulated wavepacket beam radiates superradiantly beyond the quantum cut-off condition $\omega > 1/\sigma_t$, it still has a high harmonic cut-off (Eq. 44) that depends on the tightness of the bunching.

This interesting new result suggests that all electrons in the pulse emit in phase superradiantly even if the electrons enter the interaction region <u>sparsely</u> and <u>randomly</u>, see fig.7. Because all wavepackets are modulated (by the same laser) at the same frequency and phase, they cooperatively radiate constructively at all harmonics $l\omega_b$ with coherence time equal to the electron beam pulse duration $t_{pulse}$ and corresponding spectral linewidth $\Delta\omega = 1/\sigma_{pulse}$. In a modulated wavepacket Smith-Purcell experiment as shown in Fig. 2 we expect to get a spectrum similar to the one shown in Fig. 6 but the radiated energy would be enhanced by a factor of N, and the spectral linewidth would be narrower by a factor $\sigma_{pulse}/\sigma_t$.

Another interesting observation is that the dependence on the wavepacket size disappeared altogether in (61). In fact, the emission is the same as the superradiant emission of a bunched point-particle beam [14]. Here is another expression of the wave-particle duality nature in spontaneous emission – no distinction between the spectral superradiant emission of a bunched point-particles beam and a pulse of phase correlated modulated wavepackets, even if the electron beam is tenuous and the wave packets are sparse and random (we do not rule out the possibility that the photon statistics may be different).

It is also interesting to point out that the configuration of bunching the electron wavepackets by a laser and measuring their superradiant emission at the laser modulation frequency shown in Fig. 2 is reminiscent of the Schwartz-Hora experiment [31] and its interpretation by Marcuse [32] as coherent cooperative optical transition radiation. According to Marcuse the signal to noise calculation, based on his model and the reported experimental parameters, is below the measurable level in the parameters of the experiment [32], and unfortunately there is no independent experimental confirmation of this effect. We suggest that the Smith-Purcell radiation scheme of Fig. 2 will be a more efficient radiation scheme for observing superradiant emission from a laser modulated electron wavepackets beam.

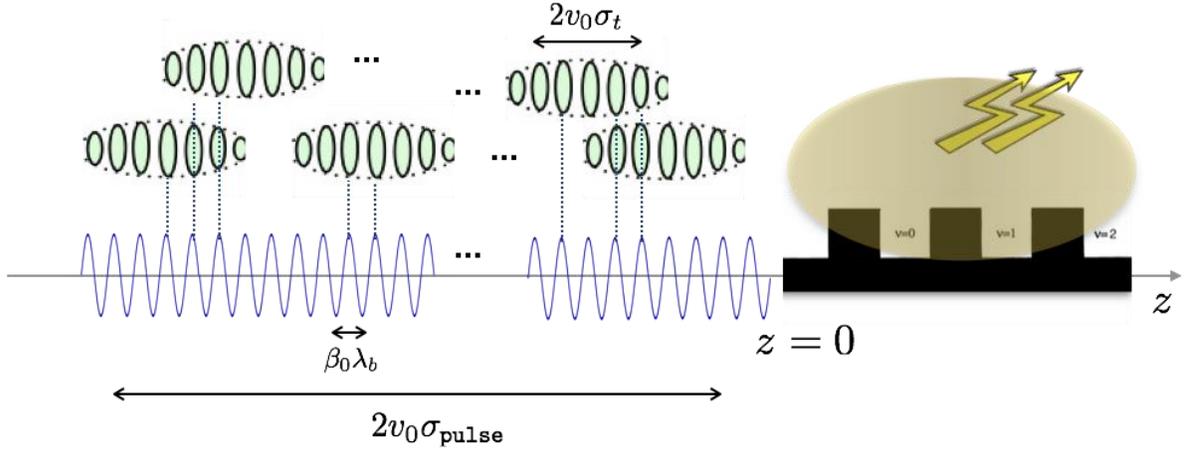

*Figure 7: A schematic diagram of superradiant coherent emission from a pulse of phase-correlated density-modulated electron quantum wavepackets entering the interaction region at random.*

**Modulated-wavepacket electron beam: Stimulated Superradiance**

In stimulated interaction (acceleration/deceleration) of a pulse of modulated wavepackets we sum up the incremental energy contributions $\left(\Delta W_{q,l,j}\right)_{e,ST-MOD}$ (eq. 37) of all modulated electron wavepackets j

$$\left(\Delta W_{q,l}\right)^{pulse}_{ST-MOD} = -eE_0 L \left|\tilde{M}_{q\perp}\right| B_l \mathrm{sinc}(\theta_0 L/2) e^{-\sigma^2_{pulse}(\omega_0 - l\omega_b)^2/2} \sum_{j=1}^{N} \cos\left[\theta L/2 + (\omega_0 - l\omega_b)t_{0j} + l\omega_b t_0 - \varphi_0\right]$$

(63A)

Replacing summation with integration over the Gaussian statistical distribution of the electrons in the pulse (see Append. D), one obtains

$$\left\langle \left(\Delta W_{q,l}\right)^{pulse}_{ST-MOD} \right\rangle = -NeE_0 L \left|\tilde{M}_{q\perp}\right| B_l \mathrm{sinc}(\theta_0 L/2) e^{-\sigma^2_{pulse}(\omega_0 - l\omega_b)^2/2} \cos\left[\theta_0 L/2 + l\omega_b t_0 - \varphi_0\right] \quad (64)$$

Resonant stimulated superradiant emission/absorption (deceleration/acceleration) takes place at synchronizm $(\theta = 0)$ if the interaction laser (see Fig. 2) is tuned to one of the beam bunching harmonic frequencies $(\omega_0 = l\omega_b)$ at proper deceleration/acceleration phase relative to the bunching - $\omega_0 t_0 - \varphi_0 = 0, \pi$. The laser frequency ($\omega_0$) resonant detuning range is determined by the duration of the electron beam pulse - $\Delta\omega < 1/\sigma_{pulse}$.

**Conclusions and Outlook**

"Does the history-dependent dimension of a free-electron quantum wavepacket have physical effect in its interaction with light? Can it be measured?" Answering this question has been the main thrust of this article. Based on the present analysis of the semi-classical (Maxwell equations) model for the radiation emission and the corresponding earlier analysis of the quantum-mechanical model for the electron wavepacket dynamics in the same set-up [27], our answer is affirmative.

In this article, we studied the spontaneous and stimulated radiation process of a single free electron wavepacket, as well as the superradiance processes in an ensemble (beam) of electron wavepackets. This analysis was carried out in the framework of a semi-classical model, in which the free electron charge density is represented by the expectation value of the probability density of wavefunction, and the radiation field is taken to be the classical field solution of Maxwell equations (solved in the framework of a mode expansion model). This work is complementary and fully consistent with our earlier analysis of stimulated interaction (acceleration/deceleration) of a single electron quantum-wavepacket, based on solution of Schrodinger equation for the electron interacting with a coherent classical (laser) field [27]. Based on the complementarity and consistency of the two independent formulations, we made the following observations, as listed in Table 1:

A. The single electron spontaneous emission and stimulated emission/absorption processes satisfy a wavepacket size-dependent cut-off frequency condition $\omega > 1/\sigma_t$ (eq. 1) (row 2 in Table 1). However, if the wavepacket is density-modulated, these radiative processes can still take place beyond the cutoff condition around harmonics of the wavepacket modulation frequency (row 2 in Table 1).

B. The consistency of the semi-classical formulas, derived independently for wavepacket-dependent radiative emission/absorption and corresponding electron deceleration/acceleration, reveals a "phase-space correspondence" conservation of energy relation between the free electron and the radiation field in the corresponding phase-space volume of the radiation mode that overlaps the electron wavepacket space and trajectory.

C. We have reveal a generalized "Einstein relation" between the spectral spontaneous radiation emitted by a free electron wavepacket into a radiation mode and its stimulated interaction energy exchange with the field of input radiation launched into the same mode. This relation can be a useful method for predicting the gain of a variety of free electron radiation schemes and devices [15], or the acceleration rate of various laser-acceleration schemes (like DLA [10]), based on measurement of spontaneous emission in the same setup configurations.

D. Our analysis of the radiative interaction of a free electron, represented by a quantum wavepacket, reveals the transition from the quantum interaction regime (e.g. as in PINEM) to the classical point-particle regime (e.g. as in DLA). This transition gives physical meaning to the quantum wavepacket function, and suggests how its size and characteristics can be measured.

Such wavepacket-dependent measurement can be based on observing the characteristic long wavelength cutoff effect for $\sigma_z(L_D) > \lambda\beta$ in a stimulated radiative interaction Smith-Purcell experiment in Fig. 1, and even more distinctly, by measuring the harmonic spectrum signature (Figs. 5,6) of a modulated wavepacket experiment (Fig. 2). It is emphasized that these suggested experimental schemes measure the 'history-dependent' size of the wavepacket $\sigma_z(t_D)$, and not its fundamental (coherence length) size $\sigma_{z_0} = \hbar/2\sigma_{p_0}$. Other schemes for measuring the quantum wavepacket characteristics that have been considered earlier, such as Compton-scattering by an electron wavepacket [24-26] or wavepacket self-interference [33,34], cannot provide such information.

It is noted that semi-classical analysis of spontaneous emission is not consistent with conventional QED theory. The semiclassical analysis produces results consistent with the classical point-particle limit theory of shot-noise spontaneous emission and superradiant emission from an electron beam [14], but it cannot produce the quantum spontaneous emission expressions that are derived in the quantum limit of an electron plane-wave function [15,24-26]. However, the semiclassical expressions of stimulated interaction of free quantum electron wavefunction are fully consistent with QED [28]. Measuring stimulated interaction of single electrons is feasible with recent significant advance in controlled generation, manipulation and modulation in real space and time of single electron quantum wavepackets [35-36, 22]. Since neither electron energy spectrum, nor radiation emission spectrum of a single electron are possible, an experiment of measuring the wavepacket dimensions requires multiple single electron experiments under the same conditions, including preselection of the electron wavepackets in space and time domains before entering the interaction region, in conditions similar to weak measurements [38].

Finally, we have also analyzed the case of spontaneous and stimulated superradiance from an ensemble (multi-particle beam) of modulated electron wavepackets, which are phase-correlated when modulated by the same laser (rows 3, 4 in Table 1). Quite interestingly, a beam of phase-correlated modulated electron wavefunctions radiates superradiantly like a classically bunched point-particles beam, even if the modulated wavepackets are injected at random and sparsely relative to the optical period. Unfortunately, in this case, the resultant radiation spectrum does not reveal anymore the individual quantum properties of the electron wavepackets.

| **Gallery: wavepacket-radiations** | Single unmodulated wavepacket | Single modulated wavepacket | Multiple unmodulated electron beam | Multiple modulated electron beam |
|---|---|---|---|---|
| Spontaneous emission | $e^{-\omega^2 \sigma_t^2}$ (eq.29) | $\|B_l\|^2 e^{-\sigma_t^2(\omega - l\omega_b)^2}$ (eq.36) | $N e^{-\omega^2 \sigma_t^2}$ (eq.53) | $N^2 B_0^2 e^{-\omega^2 \sigma_t^2}$ (eq.63) |
| Superradiance | Null | Null | $N^2 e^{-\omega^2 \sigma_{pulse}^2}$ (eq.56) | $N^2 \|B_l\|^2 e^{-\sigma_{pulse}^2(\omega - l\omega_b)^2}, l \neq 0$ (eq.63) |
| Stimulated emission/ Stimulated superradiance | $e^{-\omega_0^2 \sigma_t^2 / 2}$ (eq.29) | $\|B_l\| e^{-\sigma_t^2(\omega_0 - l\omega_b)^2 / 2}$ (eq.37) | $N e^{-\omega_0^2 \sigma_t^2 / 2}$ (eq.56) | $N \|B_l\| e^{-\sigma_{pulse}^2(\omega_0 - l\omega_b)^2 / 2}$ (eq.64) |

Table 1: *Summary of the radiative interaction frequency scaling of free quantum electron wavepackets.*

## Acknowlegements

We acknowledge A. Friedman and P. Kling for useful discussions and comments. The work was supported in parts by DIP (German-Israeli Project Cooperation) and by the PBC program of the Israel council of higher education. Correspondence and requests for materials should be addressed to A. G.(gover@eng.tau.ac.il).

## Appendix A

## Electron quantum wavepacket modulation by near-field multiphoton emission/absorption

We analyze the multiphoton emission/absorption process that takes place when a single electron quantum wavepacket traverses through the near field of a nanometric structure like a "tip" that is illuminated by an IR laser, as shown in Fig. 2. For simplicity, we assume here that for short enough interaction distances (or a tip, see fig.2) the diffraction and dispersion processes are small enough to assume that the transverse dimension $\sigma_\perp$ and the longitudinal dimension $\sigma_z = \beta_0 c \sigma_t$ stay constant throughout the near-field region, and $\sigma_t$ satisfy the uncertainty condition. Following Feist et al.[22], we model the electron wavepacket energy modulation by solving the relativistically modified Schrodinger equation[15,27] with the optical near-field perturbed Hamiltonian

$$H = H_0 - \frac{e}{m} \mathbf{p} \cdot \mathbf{A} = \frac{p^2}{2m^*} + \frac{eF(z)}{\gamma_0 m \omega_b} p \sin \omega_b t \tag{65}$$

where $m^* = \gamma_0^3 m$ is the effective mass, $\omega_b$ is the optical frequency of the modulating laser beam to modulate the wavefunction, and $F(z)$ is the slow-varying spatial distribution of the near-field. Assuming that $F(z)$ may be considered constant for all relevant momentum components of the wavepacket, the solution of the Schrodinger equation $i\hbar \frac{\partial}{\partial t} \psi = H\psi$ is expressed by Floquet expansion

$$\psi(z,t) = \sum_n c_n(t) e^{i(pz - E_p t)/\hbar} e^{in\omega_b t} \tag{66}$$

Using the Raman-Nath approximation $p^2/2m^* - E_p - n\hbar\omega_b \approx 0$, we can write the Schrodinger equation as the standard Bessel function recurrence relation

$$2\frac{\partial}{\partial \tau} c_n = c_{n-1} - c_{n+1} \tag{67}$$

which has the general solution $c_n = J_n(\tau)$ and $\tau = \dfrac{eF(z)pt}{\gamma_0 m\hbar\omega_b}$, where $J_n$ is Bessel function of order n. Thus, the wavefunction in the interaction regime is given by

$$\sum_n J_n\left(\frac{eF(z)pt}{\gamma_0 m\hbar\omega_b}\right) e^{i(pz-E_p t)/\hbar} e^{-in\omega_b t} \tag{68}$$

Then the longitudinal wavepacket after interaction is given by

$$\psi(z,t) = \left(2\pi\sigma_{p_0}^2\right)^{-1/4} \int \frac{dp}{\sqrt{2\pi\hbar}} \sum_n J_n\left(\frac{eF(z)pt}{\gamma_0 m\hbar\omega_b}\right) \exp\left(-\frac{(p-p_0)^2}{4\sigma_{p_0}^2}\right) e^{i(pz-E_p t)/\hbar} e^{in\omega_b t} \tag{69}$$

Now let us make a simple approximation for the energy gain in the interaction regime of length $L_I$, and define a parameter

$$2|g| = \left.\frac{eF(z)pt}{\gamma_0 m\hbar\omega_b}\right|_{t=L_I/v_0} \simeq \int_0^{L_I} \left(\frac{eF(z)}{\hbar\omega_b}\right) dz$$

where $2|g|$ is the averaged photon number gain from the near-field. Thus, the wavefunction after passing through the interaction regime can be expressed as

$$\psi(z,t) = \left(2\pi\sigma_{p_0}^2\right)^{-1/4} \int \frac{dp}{\sqrt{2\pi\hbar}} \sum_n J_n(2|g|) \exp\left(-\frac{(p-p_0)^2}{4\sigma_{p_0}^2}\right) e^{i(pz-E_p t)/\hbar} e^{in\omega_b t} \tag{70}$$

To simplify the expression, we expand the energy dispersion to second order:

$$E_p = c\sqrt{m_0^2 c^2 + p^2} \approx E_{p_0} + v_0(p-p_0) + \frac{(p-p_0)^2}{2m^*} \tag{71}$$

where $E_{p_0} = \gamma_0 m_0 c^2, p_0 = \gamma_0 m_0 v_0, m^* = \gamma_0^3 m_0$ are the effective energy, momentum, and mass respectively with the Lorentz factor $\gamma_0 = 1/\sqrt{1-v_0^2/c^2}$. For short interaction length (the near field of a tip) the second order quadratic term in the energy-momentum dispersion expansion (71) is neglected. After substitution in (70) with replacing $p \to p - n\hbar\omega_b/v_0$ for each term in the summation of the modulated momentum distribution (the integrand of Eq. 70) is:

$$\psi_p = \left(2\pi\sigma_p^2\right)^{-1/4} \sum_n J_n\left(2|g|\right) \exp\left(-\frac{(p - p_0 - n\delta p)^2}{4\sigma_{p_0}^2}\right) \qquad (72)$$

where $\delta p = \hbar\omega/v_0$. For free-space drift after the modulation, we have to expand the energy-momentum dispersion relation expansion (71) to second order, and then get:

$$\begin{aligned}
\psi(z,t) &= \int \frac{dp}{\sqrt{2\pi\hbar}} \psi_p e^{i(pz - E_p t)/\hbar} \\
&= \frac{e^{i(p_0 z - \varepsilon_{p_0} t)/\hbar}}{\sqrt{2\pi\hbar}} \int dp' \psi_{p'} e^{\frac{ip'}{\hbar}\left(z - v_0 t - \frac{p'^2 t}{2m^*}\right)} \\
&= \frac{e^{i(p_0 z - \varepsilon_{p_0} t)/\hbar}}{\sqrt{2\pi\hbar}} \left(2\pi\sigma_{p_0}^2\right)^{-1/4} \sum_{n=-\infty}^{\infty} J_n\left(2|g|\right) \cdot F_n(z,t)
\end{aligned} \qquad (73)$$

wit $p' = p - p_0$, and the integral $F_n(z,t)$ is:

$$\begin{aligned}
F_n(z,t) &= \int_{-\infty}^{\infty} \exp\left(-\frac{(p' - n\delta p)^2}{4\sigma_{p_0}^2}\right) e^{\frac{ip'}{\hbar}\left(z - v_0 t - \frac{p'^2 t}{2m^*}\right)} dp' \\
&= \sqrt{\pi} \left(\frac{1}{4\sigma_{p_0}^2} + \frac{it}{2m^*\hbar}\right)^{-\frac{1}{2}} \exp\left(-\frac{\left(z - v_0 t - \frac{n\delta p t}{2m^*}\right)^2}{\frac{\hbar^2}{\sigma_{p_0}^2} + \frac{it\hbar}{m^*}}\right) \exp\left(i\left(\frac{n\delta p(z - v_0 t)}{\hbar} - \frac{n^2 \delta p^2 t}{2m^*\hbar}\right)\right)
\end{aligned}$$

where we use the integration formula

$$\int \exp\left(-\frac{x^2}{a^2} + ibx\right) dx = \sqrt{\pi} a \exp\left(-\frac{a^2 b^2}{4}\right), \quad \Re(a^2) > 0 \qquad (74)$$

Finally, after performing the momentum integral, one obtains

$$\psi(z,t) = \frac{e^{i(p_0 z - \varepsilon_{p_0} t)/\hbar}}{(2\pi\sigma_{z_0}^2)^{1/4}\sqrt{1+i\xi t}} \sum_{n=-\infty}^{\infty} J_n(2|g|) \exp\left(-\frac{\left(z - v_0 t - \frac{n\delta p t}{2m^*}\right)^2}{4\sigma_{z_0}^2(1+i\xi t)}\right) e^{i\left(\frac{n\delta p}{\hbar}\right)\left(z - v_0 t - \frac{n\delta p t}{2m^*}\right)}$$

$$= \frac{e^{i(p_0 z - \varepsilon_{p_0} t)/\hbar}}{(2\pi\sigma_{z_0}^2)^{1/4}\sqrt{1+i\xi t}} \sum_{n=-\infty}^{\infty} J_n(2|g|) \exp\left(-\frac{\left(z - v_0 t - n\bar{\lambda}_c^* \omega_b t/2\beta_0\right)^2}{4\sigma_{z_0}^2(1+i\xi t)}\right) e^{i\left(\frac{n\omega_b}{v_0}\right)\left(z - v_0 t - n\bar{\lambda}_c^* \omega_b t/2\beta_0\right)}$$

(75)

The envelopes of the different harmonics develop in space similarly to the unmodulated free drifting wavepacket (Eq. 5 in the text). Here $\sigma_{z_0} = \hbar/2\sigma_{p_0}$ is the initial wavepacket width, and $\xi = \frac{\hbar}{2m^*\sigma_{z_0}^2}$ is the drift chirping factor, $\bar{\lambda}_c^* = \frac{\hbar}{m^* c}$ is an effective Compton wavelength, and $\beta_0 = v_0/c$.

**Density micro-bunching in free space drift**

In quantum mechanics, the current density operator is given by

$$\vec{J} = -\frac{e\hbar}{2m^* i}\left(\psi^*\nabla\psi - \psi\nabla\psi^*\right) \tag{76}$$

Considering the longitudinal current $J_z$ along the z-direction (Eq. 6 in the main text), we solve the terms

$$\nabla\psi = \frac{\partial}{\partial z}\psi(z,t) = \left(\frac{ip_0}{\hbar}\right)\psi(z,t) + \frac{e^{i(p_0 z - \varepsilon_{p_0} t)/\hbar}}{\sqrt{2\pi\hbar}}\int dp'\left(\frac{ip'}{\hbar}\right)\psi_{p'} e^{\frac{ip'}{\hbar}\left(z - v_0 t - \frac{p'^2 t}{2m^*}\right)}$$

$$\nabla\psi^* = \frac{\partial}{\partial z}\psi^*(z,t) = \left(\frac{-ip_0}{\hbar}\right)\psi(z,t) + \frac{e^{-i(p_0 z - \varepsilon_{p_0} t)/\hbar}}{\sqrt{2\pi\hbar}}\int dp'\left(\frac{-ip'}{\hbar}\right)\psi_{p'} e^{\frac{-ip'}{\hbar}\left(z - v_0 t - \frac{p'^2 t}{2m^*}\right)}$$

(77)

Here we assume $p_0 \gg \langle p' \rangle$, and then we only consider the contribution of the first terms. Thus, we obtain

$$J(\zeta,t) \simeq -e v_0 |\psi(\zeta,t)|^2 \tag{78}$$

with $\zeta = z - v_0 t$. Then the density probability is given by:

$$|\psi(\zeta,t)|^2 = \frac{1}{\sqrt{2\pi}\sigma_z(t)} \sum_{n,m=-\infty}^{\infty} J_n(2|g|) J_m(2|g|) \exp\left(-\frac{\left(\zeta - \frac{n\delta p t}{2m^*}\right)^2}{4\sigma_{z_0}^2(1+i\xi t)} - \frac{\left(\zeta - \frac{m\delta p t}{2m^*}\right)^2}{4\sigma_{z_0}^2(1-i\xi t)}\right)$$
$$\times \exp\left(i\left(\frac{(n-m)\delta p}{\hbar}\right)\left(\zeta - \frac{(n+m)\delta p t}{2m^*}\right)\right) \tag{79}$$

where the spatial wavepacket size of the spreading wavepacket envelope is

$$\sigma_z(t) = \sqrt{\sigma_{z_0}^2 (1+\xi^2 t^2)}.$$

Note that in the limit of no modulation ($2|g| \to 0$ in Eq. 79) reduces to the limit of Eq. 5 in the text: a Gaussian wavepacket, exhibiting expansion and phase chirp as it drifts with time. Its density probability after drift is:

$$|\psi(z,t)|^2 = \frac{1}{\sqrt{2\pi}\sigma_z(t)} \exp\left(-\frac{(z-v_0 t)^2}{2\sigma_z^2(t)}\right) \tag{80}$$

Also note that the harmonic density modulation that is implied by Eq. 15 is a direct consequence of the nonlinearity of the energy dispersion relation (third term in Eq. 71). In its absence ($\xi t = \hbar/2m^* \sigma_{z_0}^2 \ll 1$) the energy modulation does not convert into density bunching:

$$\psi(z,t) = \frac{e^{i(p_0 z - \varepsilon_{p_0} t)/\hbar}}{(2\pi\sigma_{z_0}^2)^{1/4}} \exp\left(-\frac{(z-v_0 t)^2}{4\sigma_{z_0}^2}\right) \sum_{n=-\infty}^{\infty} J_n(2|g|) \exp\left(i\frac{n\delta p}{\hbar}(z-v_0 t)\right)$$
$$= \frac{e^{i(p_0 z - \varepsilon_{p_0} t)/\hbar}}{(2\pi\sigma_{z_0}^2)^{1/4}} \exp\left(-\frac{(z-v_0 t)^2}{4\sigma_{z_0}^2}\right) \exp\left(i|g|\sin(z-v_0 t)/\sigma_{z_0}\right) \tag{17}$$

and the density distribution is then independent of drift, similarly to Eq. 80, without density modulation or expansion, even though the wavepacket remains energy (phase) modulated.

Finally, we explain here that the choice made in the main text analysis that the wavepacket is emitted at its longitudinal waist at t=0 (Eq. 5 in the main text), or equivalently – with symmetrical momentum distribution without chirp (Eq. 3 in the main text), does not limit the generality of our analysis.

If the initial wavepacket before modulation is emitted from the electron source chirped, or acquires chirp due to transport to the modulation point, or by a controlled process by streaking technique [35-36], then the chirp acquired due to the dispersive transport after modulation will combine with this prior chirping. The two effects may add together to enhance the wavepacket widening, or with negative chirping – lead to compression of the wavepacket.

Instead of an unchirped momentum distribution wavepacket (Eq. 3 in the text) one would starts in this case with a complex Gaussian wavepacket:

$$\psi_p = \left(2\pi\sigma_{p_0}^2\right)^{-1/4} \exp\left[-\left(\frac{1}{4\sigma_{p_0}^2} - iC\right)(p-p_0)^2\right] \tag{81}$$

where C is defined as a prior chirp factor and $\frac{1}{\tilde{\sigma}_p^2(0)} = \frac{1}{\sigma_{p_0}^2} - i4C$. Then the Fourier transformation to real space (Eq. 2 in the main text) results in a modified complex Gaussian wavepacket (Eq. 5 in the text) with complex wavepacket size at time t=0:

$$\tilde{\sigma}_z(0) = \frac{\hbar}{2\tilde{\sigma}_p(0)} \tag{82}$$

and then size spreading in time is given by

$$\tilde{\sigma}_z^2(t) = \tilde{\sigma}_z^2(0) + \frac{i\hbar t}{2m^*} = \frac{\sigma_{z_0}^2}{1-i\hbar^2 C \sigma_{z_0}^2} + \frac{i\hbar t}{2m^*}$$

$$= \frac{\sigma_{z_0}^2}{1+\hbar^2 C^2 \hbar^2 \sigma_{z_0}^4} + i\left(\frac{\hbar^2 C^2 \sigma_{z_0}^4}{1+\hbar^2 C^2 \sigma_{z_0}^4} + \frac{\hbar t}{2m^*}\right) \tag{83}$$

If the prior chirp factor C is selected such that

$$\frac{\hbar^2 C^2 \sigma_{z_0}^4}{1+\hbar^2 C^2 \sigma_{z_0}^4} \equiv \frac{\hbar t_c}{2m^*} \tag{84}$$

then one obtains

$$\tilde{\sigma}_z^2(t) = \frac{\sigma_{z_0}^2}{1+\hbar^2 C^2 \sigma_{z_0}^4} + i\frac{\hbar(t+t_c)}{2m^*} \tag{85}$$

With this substitution, we could consider the prior chirp effect for density bunching, as we did in the eq.12 in the context.

## Appendix B

### Spectral energy of radiation mode – general formulation

In this section, we present the formalism employed throughout this article for analyzing the excitation of electromagnetic fields by current sources distributed along a waveguide, (channel, or wiggler). The cross-correlation function of the time dependent electric component $\mathbf{E}(r,t)$ and magnetic component $\mathbf{H}(r,t)$ is given by

$$R_{\mathbf{EM}}(z,\tau) = \int_{-\infty}^{\infty} dt \left\{ \iint \left(\mathbf{E}(r,t+\tau) \times \mathbf{H}(r,t)\right) \cdot \hat{e}_z \, dxdy \right\} \tag{86}$$

According to the Wiener-Khinchine theorem [J. Goodman, statistical optics, p.73-79, Wiley, 2000], the spectral density function of the electromagnetic signal energy $S_{\mathbf{EM}}(z,\omega)$ is the Fourier transform of the cross-correlation function, which is obtained as

$$S_{\text{EM}}(z,\omega) = \int_{-\infty}^{\infty} d\tau R_{\text{EM}}(z,\tau) e^{i\omega\tau} \tag{87}$$

Substituting the expression (86) into the spectral density function (87) results in

$$\begin{aligned}
S_{\text{EM}}(z,\omega) &= \int_{-\infty}^{\infty} d\tau \left\{ \int_{-\infty}^{\infty} dt \left\{ \iint \left( \mathbf{E}(r,t+\tau) \times \mathbf{H}(r,t) \right) \cdot \hat{e}_z \, dxdy \right\} \right\} e^{i\omega\tau} \\
&= \iint \left\{ \int_{-\infty}^{\infty} \left( \int_{-\infty}^{\infty} d\tau \mathbf{E}(r,t+\tau) e^{i\omega\tau} \right) \times \mathbf{H}(r,t) dt \right\} \cdot \hat{e}_z \, dxdy \\
&= \iint \left\{ \mathbf{E}(r,\omega) \times \left( \int_{-\infty}^{\infty} \mathbf{H}(r,t) e^{-i\omega t} dt \right) \right\} \cdot \hat{e}_z \, dxdy \\
&= \iint \left\{ \mathbf{E}(r,\omega) \times \mathbf{H}(r,-\omega) \right\} \cdot \hat{e}_z \, dxdy \\
&= \iint \left\{ \mathbf{E}(r,\omega) \times \mathbf{H}^*(r,\omega) \right\} \cdot \hat{e}_z \, dxdy
\end{aligned} \tag{88}$$

In the last step, we use the relation $\mathbf{H}(r,-\omega) = \mathbf{H}^*(r,\omega)$ because $\mathbf{H}(r,t)$ is real. Finally, the total energy carried by the electromagnetic field is calculated by integrating the spectral density $S_{\text{EM}}(z,\omega)$ over the entire frequency domain, resulting in

$$\begin{aligned}
W(z) &= \frac{1}{2\pi} \int_{-\infty}^{\infty} S_{\text{EM}}(z,\omega) d\omega = \frac{1}{2\pi} \int_{-\infty}^{\infty} \left\{ \iint \left\{ \mathbf{E}(r,\omega) \times \mathbf{H}^*(r,\omega) \right\} \cdot \hat{e}_z \, dxdy \right\} d\omega \\
&= \frac{1}{2\pi} \int_{0}^{\infty} \left\{ \iint \left\{ \mathbf{E}(r,\omega) \times \mathbf{H}^*(r,\omega) + \mathbf{E}(r,-\omega) \times \mathbf{H}^*(r,-\omega) \right\} \cdot \hat{e}_z \, dxdy \right\} d\omega \\
&= \frac{2}{\pi} \int_{0}^{\infty} \left\{ \iint \frac{1}{2} \Re \left\{ \mathbf{E}(r,\omega) \times \mathbf{H}^*(r,\omega) \right\} \cdot \hat{e}_z \, dxdy \right\} d\omega
\end{aligned} \tag{89}$$

Defining $\dfrac{dW(z)}{d\omega}$ as the spectral energy distribution of the electromagnetic field ($\omega > 0$), we identify

$$\frac{dW(z)}{d\omega} = \frac{2}{\pi} \iint \frac{1}{2} \Re \left\{ \mathbf{E}(r,\omega) \times \mathbf{H}^*(r,\omega) \right\} \cdot \hat{e}_z \, dxdy \tag{90}$$

Here we defined new expressions to separate the negative and positive frequencies

$$\mathbf{E}(r,\omega) = \int_{-\infty}^{\infty} \mathbf{E}(\mathbf{r},t) e^{i\omega t} dt = \begin{cases} \mathbf{E}(r,\omega), & \omega > 0 \\ \mathbf{E}^*(r,-\omega), & \omega < 0 \end{cases}$$

$$\mathbf{H}(r,\omega) = \int_{-\infty}^{\infty} \mathbf{H}(\mathbf{r},t) e^{i\omega t} dt = \begin{cases} \mathbf{H}(r,\omega), & \omega > 0 \\ \mathbf{H}^*(r,-\omega), & \omega < 0 \end{cases}$$

(91)

and the inverse Fourier transforms of the fields are

$$\mathbf{E}(\mathbf{r},t) = \frac{1}{2\pi} \int_{-\infty}^{\infty} \breve{\mathbf{E}}(\mathbf{r},\omega) e^{i\omega t} d\omega$$

$$\mathbf{H}(\mathbf{r},t) = \frac{1}{2\pi} \int_{-\infty}^{\infty} \breve{\mathbf{H}}(\mathbf{r},\omega) e^{i\omega t} d\omega$$

And according to the Parseval theorem, the total energy is expressed as

$$W(z) = \frac{1}{2\pi} \int_{-\infty}^{\infty} S_{EM}(z,\omega) d\omega = \int_{-\infty}^{\infty} P(z,t) dt = R_{EM}(z,0)$$

(92)

where $P(z,t) = \iint (\mathbf{E}(r,t) \times \mathbf{H}(r,t)) \cdot \hat{e}_z \, dxdy$ is the instantaneous power. The radiation field that is excited by a general current $J(r,\omega)$ is expanded in the frequency domain in terms of a set of orthogonal directional transverse modes $\{\tilde{\mathbf{E}}_q(\mathbf{r}), \tilde{\mathbf{H}}_q(\mathbf{r})\}$ that are the transversely confined homogeneous solution of the electromagnetic wave equations of free space or a source-less guiding structure

$$\{\mathbf{E}(r,\omega), \mathbf{H}(r,\omega)\} = \sum_q C_q \{\tilde{\mathbf{E}}_q(\mathbf{r}), \tilde{\mathbf{H}}_q(\mathbf{r})\}$$

For calculating axial flow of radiative energy, only transverse components of the fields need to be taken into account. Using the modal expansion formalism, we represent the fields in terms of a complete set of forward and backward propagating transverse modes $q$ propagating in the z-direction):

$$\begin{aligned}\mathrm{E}(r,\omega) &= \sum_q C_q(z,\omega)\mathrm{E}_{q\perp}(r_\perp,\omega)\exp(ik_{q_z}z) + C_{-q}(z,\omega)\mathrm{E}_{-q\perp}(r_\perp,\omega)\exp(-ik_{q_z}z) \\ \mathrm{H}(r,\omega) &= \sum_q C_q(z,\omega)\mathrm{H}_{q\perp}(r_\perp,\omega)\exp(ik_{q_z}z) + C_{-q}(z,\omega)\mathrm{H}_{-q\perp}(r_\perp,\omega)\exp(-ik_{q_z}z)\end{aligned} \quad (93)$$

where $C_q, k_{q_z}, \mathrm{E}_{q\perp}, \mathrm{H}_{q\perp}$ are the slow-varying amplitude, the wave number, and the electric and magnetic field transverse profile functions of the electromagnetic mode $q$, respectively.

The spectral radiative energy emission per mode q from $\dfrac{dW(z)}{d\omega} = \sum_q \dfrac{dW_q(\omega)}{d\omega}$ is given by (eq.13 in the context)

$$\frac{dW_q(\omega)}{d\omega} = \frac{2}{\pi} P_q \left( |C_q(z,\omega)|^2 - |C_{-q}(z,\omega)|^2 \right) \quad (94)$$

where $P_q = \dfrac{1}{2}\Re \iint \tilde{E}_q(r) \times \tilde{H}_q(r) \cdot \hat{e}_z \, dxdy$ the normalization power of mode q, and $C_q(z,\omega)$ is the slowly growing field amplitude of the radiation mode q at spectral frequency $\omega$ along its propagation direction (z). Mode "$-q$" propagates in the inverted "$z$"-direction. If the electromagnetic wave is known to propagate only in the +z-direction $\left(C_{-q}(z,\omega)=0\right)$, thus

$$\frac{dW_q(\omega)}{d\omega} = \frac{2}{\pi} P_q |C_q(z,\omega)|^2 \quad (95)$$

### Appendix C

The formulation in this paper is based on spectral (Fourier transform) analysis of finite time signals. A short derivation is necessary in order to match the spectral formulation to the single frequency formulation that is generally used in connection to stimulated interaction with a coherent laser beam such as

$$\mathrm{E}_{int} = \mathrm{E}_0 \cos(\omega_0 t + \varphi_0) \quad (96)$$

This should be matched to the spectral presentation of the axial electric field in time domain

$$E_{in}(\omega) = \int_{-\infty}^{\infty} e^{i\omega t} E_{int} dt = C_q^{in}(\omega)\varepsilon_{qz} \tag{97}$$

In order to have a finite time radiation wave we truncate the field (eq.96) into a time window $-T_{win}/2 < t < T_{win}/2$ long enough, such that the interaction of the electron wavepacket interaction takes place entirely within this time duration $T_{win}$. Thus,

$$E_{in}(\omega) = E_0 \int_{-T_{win}/2}^{T_{win}/2} e^{i\omega t} \cos(\omega_0 t + \varphi_0) dt = \frac{E_0}{2} T_{win} \left( \text{sinc} \frac{(\omega-\omega_0)T_{win}}{2} + \text{sinc} \frac{(\omega+\omega_0)T_{win}}{2} \right) \tag{98}$$

Since we consider only positive frequencies, we get $C_q^{in}(\omega)\varepsilon_{qz} = \frac{E_0 T_{win}}{2} \text{sinc} \frac{(\omega-\omega_0)T_{win}}{2}$. In the limit $T_{win} \to \infty$, this can be written as Eq.31 in the main text

$$C_q^{in}(\omega)\varepsilon_{qz} = \pi E_0 \delta(\omega-\omega_0)$$

**Appendix D**

For a large number of particles $N \gg 1$, both long and short electron beam pulses can be presented in a general way. Define the particles beam bunching factor

$$b_{pulse} = \frac{1}{N} \sum_{j=1}^{N} e^{i\omega t_{0j}} \tag{99}$$

One can replace the summation over j by integration over the temporal distribution of the particles in the beam pulse $f_{pulse}(t - t_{0,pulse})$, where $\int_{-\infty}^{\infty} f_{pulse}(t_{0j}) dt_{0j} = 1$ [30,14]. For a Gaussian pulse temporal distribution $f_{pulse}(t_{0j}) = \frac{1}{(2\pi)^{1/2} \sigma_{pulse}} e^{-t_{0j}^2/2\sigma_{pulse}^2}$:

$$\langle b_{pulse} \rangle = \int f_{pulse}(t_{0j} - t_{0,pulse}) e^{i\omega t_{0j}} dt_{0j} = e^{-\sigma_{pulse}^2 \omega^2/2} e^{i\omega t_{0,pulse}} \tag{100}$$

$$\left\langle \left| b_{pulse} \right|^2 \right\rangle = \left\langle \frac{1}{N}\sum_{j=1}^{N} e^{i\omega t_{0j}} \frac{1}{N}\sum_{k=1}^{N} e^{-i\omega t'_{0k}} \right\rangle = \frac{1}{N} + \frac{1}{N^2}\left\langle \sum_{j\neq k}^{N} e^{i\omega t_{0j}} e^{-i\omega t'_{0k}} \right\rangle$$

$$= \frac{1}{N} + \frac{N(N-1)}{N^2}\int f_{pulse}(t_{0j}-t_{0,pulse})e^{i\omega t_{0j}}dt_{0j}\cdot\int f_{pulse}(t'_{0j}-t_{0,pulse})e^{-i\omega t'_{0j}}dt'_{0j} \quad (101)$$

$$= \frac{1}{N} + \left(1-\frac{1}{N}\right)e^{-\sigma_{pulse}^2\omega^2} \approx e^{-\sigma_{pulse}^2\omega^2}$$

The derivation also works for the beam bunching factor of *l*-th order harmonic $b_{pulse}^{(l)}$ as defined in eq.59 in the maintext. With this formulation we can write together explicitly the spontaneous/superradiant spectral energy of the pulse

$$\left\langle \left(\frac{dW_q}{d\omega}\right)^{pulse}_{SP/SR} \right\rangle = W_q N^2 \, sinc^2(\theta L/2)e^{-\omega^2\sigma_t^2}\left(\frac{1}{N}+\left(1-\frac{1}{N}\right)e^{-\sigma_{pulse}^2\omega^2}\right) \quad (102)$$

$$= W_q sinc^2(\theta L/2)e^{-\omega^2\sigma_t^2}\left(N+N(N-1)e^{-\sigma_{pulse}^2\omega^2}\right)$$

Since usually $\sigma_{pulse} \gg \sigma_t$, Eq. 102 reproduces the classical particle beam expressions [14] for superradiant coherent radiation $(\propto N^2)$ in the frequency range $\omega < 1/\sigma_{pulse}$ (see eq. 56A), and incoherent spontaneous shot-noise radiation $(\propto N)$ - band-limited by the quantum wavepacket condition $\omega < 1/\sigma_t$ (eq. 53).

In the calculation of the stimulated-superradiance of a pulse of correlated modulated wavepackets we replace the averaging over particles in Eq. 63A by averaging over the pulse density distribution function

$$\left\langle \left(\Delta W_{q,l}\right)^{pulse}_{ST-MOD} \right\rangle = -eE_0 L \left|\tilde{M}_{q\perp}\right| B_l sinc(\theta_0 L/2)e^{-\sigma_t^2(\omega_0-l\omega_b)^2/2}\,\mathrm{Re}\left\{e^{i(\theta_0 L/2+l\omega_b t_0-\varphi_0)}\left\langle \sum_{j=1}^{N} e^{i(\omega_0-l\omega_b)t_{0j}} \right\rangle\right\}$$

$$= -eE_0 LN\left|\tilde{M}_{q\perp}\right| B_l sinc(\theta_0 L/2)e^{-(\sigma_t^2+\sigma_{pulse}^2)(\omega_0-l\omega_b)^2/2}\cos(\theta_0 L/2+l\omega_b t_0-\varphi_0)$$

(103)

where in the averaging we use eq. 100 with $\omega$ substituted with $\omega_0-l\omega_b$.